\documentclass[12pt]{article}
\usepackage{xcolor, graphicx}
\usepackage{amsmath}
\usepackage{caption,subcaption}
\usepackage[figuresright]{rotating}
\usepackage{longtable}
\usepackage{geometry}
\usepackage{amsfonts}
\usepackage{color}
\usepackage{tabularx}
\usepackage{multirow}
\usepackage{multicol}
\usepackage{array}
\usepackage{cite}
\usepackage{booktabs}
\usepackage{amssymb}
\usepackage{rotating}
\usepackage[toc,page]{appendix}

\allowdisplaybreaks[4]

\renewcommand{\theequation}{\arabic{section}.\arabic{equation}}
\def\s{\sigma}

\def\V={{{\bf\rm{V}}}}

\def\Cb{\mathbb{C}}

\def\beq{\begin{equation}}
\def\eeq{\end{equation}}
\def\bea{\begin{eqnarray}}
\def\eea{\end{eqnarray}}
\def\ba{\begin{array}}
\def\ea{\end{array}}
\def\no{\nonumber}

\def\lt{\left}
\def\rt{\right}

\title{Thermodynamic limit of the spin-$\frac{1}{2}$ XYZ spin chain with the antiperiodic boundary condition}
\author{Zhirong Xin${}^{a}$, Yusong Cao${}^{b,c}$, Xiaotian Xu${}^{b}\footnote{Corresponding author:
xtxu@nwu.edu.cn}$, Tao Yang${}^{d,e,f}$, \\Junpeng Cao${}^{b,c,f,g}$ and Wen-Li Yang${}^{d,e,f}$}

\begin{document}
\date{}
\maketitle
\begin{center}
${}^a$ School of Physics and Electronic Information, Baicheng Normal University, China\\
${}^b$ Beijing National Laboratory for Condensed Matter Physics, Institute of Physics, Chinese Academy of Sciences, Beijing 100190, China\\
${}^c$ Songshan Lake Materials Laboratory, Dongguan, Guangdong 523808, China\\
${}^d$ Institute of Modern Physics, Northwest University, Xi’an 710127, China\\
${}^e$ Shaanxi Key Laboratory for Theoretical Physics Frontiers, Xian 710069, China\\
${}^f$ Peng Huanwu Center for Fundamental Theory, Xian 710127, China\\
${}^g$ School of Physical Sciences, University of Chinese Academy of Sciences, Beijing, China\\
E-mail: zhirongxin@foxmail.com, caoyusong15@mails.ucas.an.cn, xtxu@nwu.edu.cn, yangt@nwu.edu.cn, junpengcao@iphy.ac.cn, wlyang@nwu.edu.cn

\end{center}

\begin{abstract}
Based on its off-diagonal Bethe ansatz solution, we study the thermodynamic limit of the spin-$\frac{1}{2}$ XYZ spin chain with the antiperiodic boundary condition. The key point of our method is that there exist some degenerate points of the crossing parameter $\eta_{m,l}$, at which the associated inhomogeneous $T-Q$ relation becomes a homogeneous one.
This makes extrapolating the formulae deriving from the homogeneous one to an arbitrary $\eta$ with  $O(N^{-2})$ corrections for a large $N$ possible.
The ground state energy and elementary excitations of the system are obtained.
By taking the trigonometric limit, we also give the results of antiperiodic XXZ spin chain within the gapless region in the thermodynamic limit, which does not have any degenerate points.

\vspace{0.5truecm}

\end{abstract}

\section{Introduction}

The spin-$\frac{1}{2}$ XYZ chain is a typical quantum integrable model and
has many applications in statistical mechanics, quantum magnetism, string theory and mathematical physics \cite{0,Korepin1993,00}.
The periodic XYZ spin chain (or equivalent to the eight vertex model) with even number of sites  was solved by
Baxter \cite{baxter1971phys1,baxter1971phys2,baxter2000partition, baxter1972one}.
By using the generalized algebraic Bethe ansatz \cite{takhtadzhan1979quantum},
Takhtajan and Faddeev recovered the Baxter's solution, where a suitable vacuum state is required. However, the
proper vacuum state can be constructed only for the even number of sites case.
After that, many interesting works have been done based on the exact solutions, please see \cite{takahashi2005thermodynamics} and references therein.
In 2013, a more general method named off-diagonal Bethe ansatz (ODBA) was proposed \cite{cao2013off} and the
exact solution of the XYZ spin chain with arbitrary number of sites is obtained \cite{cao2014spin,wang2015off},
where the eigenvalue of transfer matrix is given by an inhomogeneous $T-Q$ relation.
Besides the periodic boundary condition, the XYZ model with antiperiodic and integrable open ones have been  studied extensively \cite{Inami1994,Houboyu1995,Fan96,zhang1}.

So far many attempts have been done to solve the resulting BAEs from inhomogeneous $T-Q$ relations \cite{Nep13, li2014thermodynamic, Wen18, xin2018thermodynamic, qiao2018twisted,Sun19}, the corresponding distribution of Bethe roots for ground-state or elementary excitation states  is still an interesting open problem \cite{wang2015off}.
Here we propose a way to study the thermodynamic limit of the spin-$\frac{1}{2}$ XYZ chain with an antiperiodic boundary condition, which was successfully applied to the XXZ spin chain with open boundary conditions \cite{li2014thermodynamic}.
The main idea is that there exist some degenerate points of the crossing parameter (such as (\ref{Constraint-1}) below),  at which the associated inhomogeneous $T-Q$ relation
is reduced to the conventional/homogeneous one. This allows us to use  the standard method to study the thermodynamic limit \cite{Yangthermo1,Yangthermo2,Yangthermo3,Yangthermo4,takahashi1972one}.
In the thermodynamic limit, the degenerate points become dense in the whole complex plain. Thus the
exact results at degenerate points could approach to the actual values of physical quantities.

However, for some quantum integrable models such as the
XXZ spin chain with antiperiodic boundary condition \cite{1,2,3,4},
there is no the degenerate point and the inhomogeneous
BAEs can not be reduced to the traditional product ones. It is well-known that the antiperiodic XXZ spin chain is an important model and
many interesting phenomena such as edge states, zero modes, boundary bound states and topological excitations are induced by the twisted boundaries \cite{xin2018thermodynamic,qiao2018twisted,qiao}.
Thus the study of the antiperiodic XXZ spin chain is another motivation of this work.
The difficulty of lacking degenerate points can be overcome if we consider a more general model such as the XYZ case (the antiperiodic XXZ chain is its trigonometric limit) which has some degenerate points allowing us to
investigate its thermodynamic limit.

In this paper, we study the thermodynamic limit of the antiperiodic XYZ spin chain. We obtain the distribution of Bethe roots, string
structures, ground state energy and typical elementary excitations. We also check these results by the numerical calculations and the finite size scaling analysis.
Particularly, we obtain the exact results of the antiperiodic XXZ spin chain in the thermodynamic limit.
In order to see the boundary effects clearly, we also give the results of
the XYZ spin chain with the periodic boundary condition as the comparisons.

The paper is organized as follows. In the next section, the model Hamiltonian and exact solutions are introduced.
In section 3, we study the thermodynamic limit of the antiperiodic XYZ model at degenerate points.
In section 4, we generalize these results to the system with general interactions.
As an important component, the results of the antiperiodic XXZ model are given in section 5.
Section 6 is the concluding remarks and discussions. Some supporting materials are given in Appendices A--C.

\section{Antiperiodic XYZ model and its exact solutions}

\setcounter{equation}{0}
Let us fix a generic complex number $\eta$ and a generic imaginary number $\tau$ such that ${\rm
Im}(\tau)>0$, which related to the coupling constants. The most anisotropic spin-$\frac{1}{2}$ XYZ chain is described by the Hamiltonian
\begin{eqnarray}\label{Hamiltoniant}
H=\frac{1}{2} \sum^N_{j=1} \left[ J_x(\eta,\tau) \sigma^x_j \sigma^x_{j+1} + J_y(\eta,\tau) \sigma^y_j \sigma^y_{j+1} + J_z(\eta,\tau) \sigma^z_j \sigma^z_{j+1} \right],\label{Ham-XYZ}
\end{eqnarray}
where $J_x (\eta,\tau)$, $J_y(\eta,\tau)$ and $J_z(\eta,\tau)$ are the anisotropic coupling constants
\begin{eqnarray}\label{Jf}
  J_x(\eta,\tau)= \frac{\theta_{01}(\eta)}{\theta_{01}(0)}, \quad J_y (\eta,\tau)= \frac{\theta_{00}(\eta)}{\theta_{00}(0)},
  \quad J_z(\eta,\tau)= \frac{\theta_{10}(\eta)}{\theta_{10}(0)},
\end{eqnarray}
$\theta_{00}(u)$, $\theta_{01}(u)$ and $\theta_{10}(u)$ are the elliptic theta functions defined in Appendix A, $\eta$ and $\tau$ are two model parameters,
and $\sigma^{\alpha}_j \ (\alpha =x,y,z)$ are the Pauli matrices. Here we consider the Hamiltonian with the antiperiodic boundary condition
\begin{eqnarray}\label{antic}
  \sigma^{x}_{N+1} =\sigma^x_1\,\sigma^x_1\,\sigma^x_1=\sigma^x_1,\quad \sigma^{y}_{N+1} =\sigma^x_1\,\sigma^y_1\,\sigma^x_1=-\sigma^y_1,\quad \sigma^{z}_{N+1} =\sigma^x_1\,\sigma^z_1\,\sigma^x_1=-\sigma^z_1.\label{Antiperiodic}
\end{eqnarray}
It is well-known that the XYZ chain with the antiperiodic boundary condition given by (\ref{Ham-XYZ}) and (\ref{Antiperiodic}) is integrable, which is guaranteed by the eight-vertex $R$-matrix $R(u)\in {\rm End}(\mathbb{C}^2\otimes \mathbb{C}^2)$ given by \cite{0}
\begin{eqnarray}
R(u)=\left(\begin{array}{llll}\alpha(u)&&&\delta(u)\\&\beta(u)&\gamma(u)&\\
&\gamma(u)&\beta(u)&\\ \delta(u)&&&\alpha(u)\end{array}\right),
\label{r-matrix}
\end{eqnarray} with the non-zero entries
\begin{eqnarray}
&&\hspace{-2.0truecm}\alpha(u)\hspace{-0.1truecm}=
 \hspace{-0.1truecm}\frac{\theta\left[\begin{array}{c} 0\\\frac{1}{2}
 \end{array}\right]\hspace{-0.16truecm}(u,2\tau)\hspace{0.12truecm}
 \theta\left[\begin{array}{c} \frac{1}{2}\\[2pt]\frac{1}{2}
 \end{array}\right]\hspace{-0.16truecm}(u+\eta,2\tau)}{\theta\left[\begin{array}{c} 0\\\frac{1}{2}
 \end{array}\right]\hspace{-0.16truecm}(0,2\tau)\hspace{0.12truecm}
 \theta\left[\begin{array}{c} \frac{1}{2}\\[2pt]\frac{1}{2}
 \end{array}\right]\hspace{-0.16truecm}(\eta,2\tau)},\quad
\beta(u)\hspace{-0.1truecm}=\hspace{-0.1truecm}\frac{\theta\left[\begin{array}{c}
 \frac{1}{2}\\[2pt]\frac{1}{2}
 \end{array}\right]\hspace{-0.16truecm}(u,2\tau)\hspace{0.12truecm}
 \theta\left[\begin{array}{c} 0\\\frac{1}{2}
 \end{array}\right]\hspace{-0.16truecm}(u+\eta,2\tau)}
 {\theta\left[\begin{array}{c} 0\\\frac{1}{2}
 \end{array}\right]\hspace{-0.16truecm}(0,2\tau)\hspace{0.12truecm}
 \theta\left[\begin{array}{c} \frac{1}{2}\\[2pt]\frac{1}{2}
 \end{array}\right]\hspace{-0.16truecm}(\eta,2\tau)},\no\\[6pt]
&&\hspace{-2.0truecm}\gamma(u)\hspace{-0.1truecm}=
 \hspace{-0.1truecm}\frac{\theta\left[\begin{array}{c} 0\\\frac{1}{2}
 \end{array}\right]\hspace{-0.16truecm}(u,2\tau)\hspace{0.12truecm}
 \theta\left[\begin{array}{c} 0\\\frac{1}{2}
 \end{array}\right]\hspace{-0.16truecm}(u+\eta,2\tau)}
 {\theta\left[\begin{array}{c} 0\\\frac{1}{2}
 \end{array}\right]\hspace{-0.16truecm}(0,2\tau)\hspace{0.12truecm}
 \theta\left[\begin{array}{c} 0\\\frac{1}{2}
 \end{array}\right]\hspace{-0.16truecm}(\eta,2\tau)},\quad
\delta(u)\hspace{-0.1truecm}=\hspace{-0.1truecm}\frac{\theta\left[\begin{array}{c}
 \frac{1}{2}\\[2pt]\frac{1}{2}
 \end{array}\right]\hspace{-0.16truecm}(u,2\tau)\hspace{0.12truecm}
 \theta\left[\begin{array}{c} \frac{1}{2}\\[2pt]\frac{1}{2}
 \end{array}\right]\hspace{-0.16truecm}(u+\eta,2\tau)}
 {\theta\left[\begin{array}{c} 0\\\frac{1}{2}
 \end{array}\right]\hspace{-0.16truecm}(0,2\tau)\hspace{0.12truecm}
 \theta\left[\begin{array}{c} 0\\\frac{1}{2}
 \end{array}\right]\hspace{-0.16truecm}(\eta,2\tau)},\label{r-func}
\end{eqnarray}
where the associated elliptic functions are defined in Appendix A. In addition to satisfying the quantum Yang-Baxter equation (QYBE),
\begin{eqnarray}
\hspace{-1.2truecm}R_{12}(u_1-u_2)R_{13}(u_1-u_3)R_{23}(u_2-u_3)=R_{23}(u_2-u_3)R_{13}(u_1-u_3)R_{12}(u_1-u_2),\label{QYB}
\end{eqnarray}
the $R$-matrix also possesses the $Z_2$-symmetry
\bea
\s^i_1\s^i_2R_{12}(u)=R_{12}(u)\s^i_1\s^i_2,\quad
\mbox{for}\,\,
i=x,y,z.\label{Z2-sym}
\eea

Throughout this paper we adopt the standard notations:
for any matrix $A\in {\rm End}(\Cb^2)$, $A_j$ is an embedding
operator in the tensor space $\Cb^2\otimes \Cb^2\otimes\cdots$,
which acts as $A$ on the $j$-th space and as identity on the other
factor spaces; $R_{i\,j}(u)$ is an embedding operator of $R$-matrix
in the tensor space, which acts as identity on the factor spaces
except for the $i$-th and $j$-th ones. Let us introduce the monodromy matrix
\begin{eqnarray}
T_0(u)&=&R_{0N}(u)\ldots R_{01}(u).\label{Mon-1}
\end{eqnarray}
The transfer matrix $t(u)$ of the XYZ chain with the antiperiodic boundary
condition (\ref{Antiperiodic}) is given by
 \begin{eqnarray}
 t(u)=tr_0\lt\{\sigma^x_0\,T_0(u)\rt\},\label{trans-per}
 \end{eqnarray} where $tr_0$ denotes the trace over the
``auxiliary space" $0$. The Hamiltonian (\ref{Ham-XYZ}) with the antiperiodic boundary condition (\ref{Antiperiodic}) can be given in terms of the transfer matrix (\ref{trans-per})
\begin{eqnarray}
H=\frac{\sigma(\eta)}{\sigma'(0)}\left\{\frac{\partial \ln
t(u)}{\partial
u}|_{u=0}-\frac{1}{2}N\zeta(\eta)\right\},\label{ham}
\end{eqnarray}
where $\sigma(u)=\theta_{11}(u,\tau)$, $\sigma'(0)=\left.\frac{\partial}{\partial
u}\,\sigma(u)\right|_{u=0}$ and the function $\zeta(u)=\frac{\partial}{\partial
u}\,\ln\sigma(u)$. The QYBE
(\ref{QYB}) and the $Z_2$-symmetry (\ref{Z2-sym})  lead to that the transfer matrices with different
spectral parameters are mutually commutative \cite{0, Korepin1993}, i.e.,
$[t(u),t(v)]=0$, which guarantees the integrability of the model by
treating $t(u)$ as the generating functional of the conserved
quantities.

The eigenvalue of the transfer matrix $t(u)$, denoted by $\Lambda(u)$, is given by an inhomogeneous $T-Q$ relation \cite{cao2014spin}
\begin{eqnarray}
\Lambda(u)&=&e^{\{i\pi(2l_1+1)u+i\phi\}}\frac{\sigma^N(u+\eta)}{\sigma^N(\eta)}\frac{Q_1(u-\eta)}{Q_2(u)}\no\\[6pt]
&&-\frac{e^{-i\pi(2l_1+1)(u+\eta)-i\phi}\sigma^N(u)}{\sigma^N(\eta)}\frac{Q_2(u+\eta)}{Q_1(u)}\nonumber\\[6pt]
&&+\frac{c\,e^{i\pi u}\sigma^{L_1}(u+\frac{\eta}{2})\sigma^N(u+\eta)\sigma^N(u)}{Q_1(u)Q_2(u)\sigma^N(\eta)\sigma^N(\eta)},\label{T-Q-Main-Anti}
\end{eqnarray}
where $l_1$ is a certain integer and $L_1$ ia a non-negative integer such that $N+L_1=2M$,  the $Q$-functions $Q_1(u)$, $Q_2(u)$ are some elliptic polynomials of degree $M$
\begin{eqnarray}
Q_1(u)&=&\prod_{j=1}^{M}\frac{\sigma(u-\mu_j)}{\sigma(\eta)},\,\,
Q_2(u)=\prod_{j=1}^{M}\frac{\sigma(u-\nu_j)}{\sigma(\eta)}.\label{Q-1}
\end{eqnarray}
The $2M+2$ parameters $\{ \mu_j \}$, $\{ \nu_j \}$, $c$ and $\phi$ should satisfy the associated BAEs
\begin{eqnarray}
  && (\frac{N}{2}-M)\eta -\sum^M_{j=1}(\mu_j-\nu_j) = (l_1 +\frac{1}{2})\tau +m_1, \quad l_1,m_1 \in \textbf{Z},  \nonumber  \\
  && M\eta +\sum^M_{j=1}(\mu_j +\nu_j) = \frac{1}{2}\tau +m_2, \quad m_2\in \textbf{Z},  \nonumber \\
  &&  c e^{[ 2i\pi(l_1+1)\mu_j +2i\pi(l_1+\frac{1}{2} )\eta +i\phi ]}\sigma^{L_1}(\mu_j+\frac{\eta}{2}) \sigma^N(\mu_j +\eta) = \prod^M_{l=1}\sigma(\mu_j-\nu_l) \sigma(\mu_j-\nu_l+ \eta),  \nonumber \\
  && c e^{-2i\pi l_1 \nu_j -i\phi } \sigma^{L_1} (\nu_j +\frac{\eta}{2}) \sigma^N(\nu_j) =- \prod^M_{l=1}\sigma(\nu_j-\mu_l) \sigma(\nu_j-\mu_l- \eta) ,\nonumber \\
  &&  e^{i\phi}\prod^M_{j=1} \frac{\sigma(\mu_j+\eta)}{\sigma(\nu_j)}=e^{\frac{i\pi k_1}{N}}, \quad k_1 = 1,\cdots, 2N, \label{bat2}
\end{eqnarray}
The eigenvalues of Hamiltonian (\ref{Hamiltoniant}) with the antiperiodic boundary condition (\ref{Antiperiodic}) is then given in terms of the Bethe roots\footnote{It is remarked that for any choice of $l_1$, $m_1$, $m_2$ and $L_1$ the solutions to (\ref{bat2}) may give rise to the complete set of the eigenvalues of the transfer matrix $t(u)$, which is a conjecture based on the numerical evidence\cite{wang2015off, cao2014spin}.} as \cite{wang2015off, cao2014spin}
\begin{eqnarray}\label{eigE01}
  E(\eta, \tau)=\frac{\sigma(\eta)}{\sigma'(0)} \left\{ \sum^M_{j=1} \left[ \frac{\sigma'(\nu_j)}{\sigma(\nu_j)}
  -\frac{\sigma'(\mu_j +\eta)}{\sigma(\mu_j +\eta)} \right] +\frac{N}{2}\frac{\sigma'(\eta)}{\sigma(\eta)} + i\pi(2l_1+1) \right\},
\end{eqnarray}
where  $\sigma'(u)=\frac{\partial}{\partial u} \sigma(u)$.

Although many attempts have been done to solve the resulting BAEs from inhomogeneous $T-Q$ relations \cite{Nep13, li2014thermodynamic, Wen18, xin2018thermodynamic, qiao2018twisted,Sun19}, the corresponding distributions of Bethe roots for ground-state or elementary excitation states  is still an interesting open problem.
This motivates us in this paper
to look for another way, instead of solving the BAEs (\ref{bat2}) for a large $N$, to study  the thermodynamic limit of the spin-$\frac{1}{2}$ XYZ chain with the antiperiodic boundary condition.

\section{Thermodynamic limit at the degenerate points}
\setcounter{equation}{0}
\label{sec-Thermo}

%\subsection{Degenerate points}

It was shown \cite{wang2015off, cao2014spin} that if the crossing parameter $\eta$ takes the
discrete values
\begin{eqnarray}\label{eta}
  \eta_{m_1,l_1}=\frac{2l_1+1}{N-2 M}\tau +\frac{2m_1}{N-2 {M}}, \quad l_1, m_1\in \textbf{Z},\label{Constraint-1}
\end{eqnarray}
there exists a solution  with $c=0$ of Eq.(\ref{bat2}), and the resulting $T-Q$ relation reduces to the conventional/homogeneous one
\begin{eqnarray}
\Lambda(u)&=&e^{\{i\pi(2l_1+1)u+i\phi\}}\frac{\sigma^N(u+\eta)}{\sigma^N(\eta)}\frac{Q(u-\eta)}{Q(u)}\no\\[6pt]
&&-\frac{e^{-i\pi(2l_1+1)(u+\eta)-i\phi}\sigma^N(u)}{\sigma^N(\eta)}\frac{Q(u+\eta)}{Q(u)},\no
\end{eqnarray}
where $Q(u)$ is an elliptic polynomial with a degree of $M$
\bea
Q(u)=\prod_{j=1}^M\frac{\s(u-\lambda_j)}{\s(\eta)}. \label{T-Q-Hom-1}
\eea
Without losing the generality\footnote{It is straightforward to generalize our method in this paper to study the most general case (\ref{eta}).}, we  take $M=N$, $l_1=-1$, $m_1=-m$ as an example to demonstrate our method. In this particular case,  Eq.(\ref{eta}) becomes
\begin{eqnarray}\label{eta2}
  \eta_{-m,-1} =\frac{\tau}{N} + \frac{2m}{N}\equiv \eta_m, \quad m\in \textbf{Z}.
\end{eqnarray}
It is clear that the degenerate point $\eta_m$ (\ref{eta2}) contains an imaginary part $\frac{\tau}{N}$.
For a finite $\tau$, the imaginary part of $\eta_m$ will tend to zero in the thermodynamic limit $N \rightarrow \infty$.
Further, we require\footnote{For the case of $0<\eta\leq\frac{1}{2}$, we can obtain
the distributions of the Bethe roots for the ground state and hole excitation in the following parts of the paper. The generalization to the case of $\frac{1}{2}<\eta\leq 1$ is straightforward although the corresponding distributions are slightly different.  Hence without losing the generality we restrict ourselves in the region $0<\eta\leq\frac{1}{2}$ in this paper. }  $0<\frac{2m}{N} \leq \frac{1}{2}$ in this paper. At the degenerate point of $\eta_m$, the BAEs (\ref{bat2}) reduce to
\begin{eqnarray}
  && e^{\pi x_j+2i\phi} \frac{\sigma^N [ \frac{i}{2}(x_j-\eta_m i) ] }{\sigma^N [\frac{i}{2}(x_j+\eta_m i) ]}
  =-\prod^N_{k\neq j} \frac{\sigma [\frac{i}{2}(x_j-x_k -2\eta_m i)] }{\sigma [\frac{i}{2}(x_j-x_k +2\eta_m i)] },\quad j=1,\cdots, N, \label{BAE22-1} \\
  &&e^{i\phi} \prod^N_{j=1} \frac{\sigma [\frac{i}{2}(x_j-\eta_m i)] }{\sigma [\frac{i}{2}(x_j +\eta_m i)] }=e^{\frac{i\pi k_1}{N}}, \quad k_1 = 1,\cdots, 2N, \label{BAE22}
\end{eqnarray}
where $\{ x_j \}$ are the Bethe roots related to the parameters in (\ref{T-Q-Hom-1}) by $\lambda_j=\frac{i}{2}x_j-\frac{\eta_m}{2}$. The BAEs (\ref{BAE22-1}) and (\ref{BAE22}) can give the complete set of solutions of the Hamiltonian \cite{cao2014spin,54}.
The eigenvalue of the Hamiltonian (\ref{Hamiltoniant}) with a fixed $\eta_{m}$ given by (\ref{eta2}) is expressed in terms of the Bethe roots as
\begin{eqnarray}\label{eigE2}
  E(\eta_m)=\frac{\sigma(\eta_m)}{\sigma'(0)} \left\{ \sum^N_{j=1} \left[ \frac{\sigma' [\frac{i}{2}(x_j +\eta_m i)]}
  {\sigma [\frac{i}{2}(x_j +\eta_m i)]} -\frac{\sigma'[\frac{i}{2}(x_j -\eta_m i)]}{\sigma [\frac{i}{2}(x_j -\eta_m i)]} \right] +\frac{N}{2}\frac{\sigma'(\eta_m)}{\sigma(\eta_m)} -i\pi \right\}.
\end{eqnarray}

\subsection{String hypothesis}

Takahashi proposed that the general solutions of Bethe ansatz equations are
\cite{takahashi1972one, takahashi2005thermodynamics, Hida1981}
\begin{eqnarray}\label{str}
  x^{j, k}_{\alpha}=x^{j}_{\alpha} +(n_j+1 -2k)\eta i +\frac{1-v_j}{2} i +O(e^{-\delta N}), \quad \quad 1\leq k \leq n_j,
\end{eqnarray}
where $x^{j}_{\alpha}$ is the position of the $j$-string on the real axis, $k$ means the $k$th Bethe roots in $j$-string, $n_j$ is the length of $j$-string,
$O(e^{-\delta N})$ means the finite size correction, and
$v_j=\pm1$ denotes the parity of $j$-string.
The center of $j$-string is the real axis if $v_j=1$, while
the center of $j$-string is the line with fixed imaginary part $i$ in the complex plane if $v_j=-1$.
Eq.(\ref{str}) is the famous string hypothesis.

Similarly as the string hypothesis (\ref{str}), we  assume that there might exist the string solutions of the type (\ref{str}) for
the Bethe ansatz equations (\ref{BAE22-1})-(\ref{BAE22}) with a large $N$.
In our case, the length $n_j$ and parity $v_j$ of $j$-string are uniquely determined by the crossing parameter $\eta$. Focusing on the
degenerate points $\eta_m$, the string structures are uniquely determined by the $\frac{2m}{N}$.
In order to show the string structures more clearly, we expand the $\frac{2m}{N}$ into a simple continued fraction (SCF) with length $l$ as that of \cite{takahashi2005thermodynamics}
\begin{eqnarray}\label{scf}
\frac{2m}{N}=\frac{c_2}{c_1}=\cfrac{1}{a_1+\cfrac{1}{\cdots +\cfrac{1}{ a_l }}},\quad a_l \geq2.
\end{eqnarray}
Here $c_1$ and $c_2$ are co-prime numbers, and $a_1\geq 2$ because of $0< \frac{2m}{N} \leq \frac{1}{2}$. For convenience, we rewrite the above SCF (\ref{scf}) as a compact form
\begin{eqnarray}
  \frac{2m}{N}=\frac{c_2}{c_1} =[a_1,a_2,\cdots,a_l], \quad a_l \geq 2.\label{ytao}
\end{eqnarray}
Define the integers $z_s$ and quantities $y_s$ as
\begin{eqnarray}
  && z_0=0, \quad z_k=\sum^k_{j=1} a_j, \quad k= 1,2,\cdots, l, \nonumber \\
  && y_{-1}=0, \quad y_0=1, \quad y_k =a_k y_{k-1} + y_{k-2}. \label{yf}
\end{eqnarray}
Then the length $n_j$ and parity $v_j$ of string solutions should satisfy \cite{takahashi1972one, takahashi2005thermodynamics}:
\begin{eqnarray}
  &&  n_j = y_{s-1} +(j-z_s)y_s,
   \quad s=0, 1, \cdots, l,  \quad z_s \leq j < z_{s+1}, \quad   j=1,2,\cdots , z_l,
   \nonumber\\
  && n_{z_l+1} = y_l, \nonumber\\
  && v_j =(-1)^{\lfloor (n_j-1)\frac{2m}{N} \rfloor}, \quad j\neq z_1,
  \nonumber\\
  && v_{z_1} = -1. \label{str-rule}
\end{eqnarray}
Here $\lfloor{x} \rfloor$ denotes the maximum integer less than or equal to $x$ (the Gauss symbol).
From the Eqs.(\ref{scf}) and (\ref{yf}), we have $y_l=c_1$ which corresponds to the length of the $(z_l+1)$-string.
From Eq.(\ref{str-rule}), we see that the number of string types is $z_l+1$.
We note that the $(z_l+1)$-string with length $n_{z_l+1}=y_l$ should be considered in the present case, because
the corresponding energy is not zero.
\begin{table}[!htp]
  \centering
\caption{Length $n_j$ and parity $v_j$ of strings for a given $\frac{2m}{N}=\frac{10}{73}=[7,3,3]$.
  From Eq.(\ref{ytao}), we have $a_1=7$, $a_2=3$ and $a_3=3$. The values of $n_j$ and $v_j$ for all the strings are
  determined by Eqs.(\ref{yf})-(\ref{str-rule}). We also list the values of $q_j$ [please see Eq.(\ref{pf}) below] for late use. }
  \begin{tabular}{|c|cc c c c c c c c c c c c c c|}
     \hline
     $j$   & & $1$ & $2$ & $3$ & $4$ & $5$ & $6$ & $7$ & $8$ & $9$ & $10$ & $11$ & $12$ & $13$ & $14$ \\ \hline
     $n_j$ & & $1$ & $2$ & $3$ & $4$ & $5$ & $6$ & $1$ & $8$ & $15$ & $7$ & $29$ & $51$ & $22$ & $73$ \\  %\hline
     $v_j$ & & $+$ & $+$ & $+$ & $+$ & $+$ & $+$ & $-$ & $+$ & $-$  & $+$ & $-$  & $+$  & $+$  &  $-$ \\
     $q_j$ & & $\frac{63}{10}$ & $\frac{53}{10}$ & $\frac{43}{10}$ & $\frac{33}{10}$ & $\frac{23}{10}$ & $\frac{13}{10}$ & $-1$ & $-\frac{7}{10}$ & $-\frac{4}{10}$
     & $\frac{3}{10}$ & $\frac{2}{10}$  & $\frac{1}{10}$  & $-\frac{1}{10}$  &  $0$ \\
     $z_s$ &{\footnotesize $z_0=0$ } & &  &  &  &  &  & {\footnotesize $z_1=7$ }&  &
     & {\footnotesize $z_2=10$} &   &  & {\footnotesize $z_3=13$}  &   \\
     \hline
   \end{tabular}
  \label{table}
\end{table}

\begin{figure}[!htp]
 \centering
    \includegraphics[height=6cm]{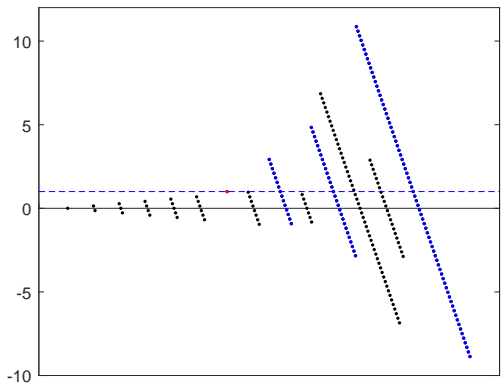}
\caption{
Strings on the complex plane for $N=73$, $\tau=i$ and $\frac{2m}{N}=\frac{10}{73}$.
The strings are arranged from left to right in the order of string length determined by Eq.(\ref{str-rule}).
The centers of strings with parity $v_j=1$ are on the real axis, while the centers of strings with parity $v_j=-1$ are on the blue dashed line which
is given by shifting the real axis up by one.
All the strings are not parallel to the vertical axis, because the degenerate crossing parameters $\eta_m$ contain an imaginary part with order $N^{-1}$.
}\label{fig4}
\end{figure}

In order to shown the string structure clearly, we give an example $\frac{2m}{N}=\frac{10}{73}=[7,3,3]$ in Table \ref{table}. From it,
the length $n_j$ and parity $v_j$ of strings can be found.
Taking the data in Table \ref{table} into Eq.(\ref{str}), we can visualize the shape of the strings in the complex plane.
The results with $N=73$, $\tau=i$ and $\eta_m=\frac{i}{73} + \frac{10}{73}$ are shown in Fig.\ref{fig4}.
From it we see that the string solutions are not parallel to the imaginary axis any more, this is because that
the parameters $\eta_m$ (\ref{eta2}) has an imaginary part with order $N^{-1}$.

\subsection{Distribution of Bethe roots}

Substituting Eq.(\ref{str}) into (\ref{BAE22-1}) and omitting the exponentially small corrections,
and then taking the product of the BAEs (\ref{BAE22}) for $n_j$ components of a $j$-string, we obtain following equation for the position $x^j_{\alpha}$ of $j$-string
\begin{eqnarray}\label{BAE3}
&&    e^{\varphi_j(x^j_{\alpha})} g^N(x^j_{\alpha}; n_j,v_j)=(-1)^{n_j+1} \prod^{z_l+1}_{r=1} \prod^{M_r}_{\beta=1} \prod^{\text{min}(n_r, n_j)-1}_{k=1}
g(x^j_{\alpha}-x^r_{\beta};  n_r+n_j ,v_r v_j)  \nonumber\\
  && \qquad \times g(x^j_{\alpha}-x^r_{\beta};  |n_r-n_j|, v_r v_j) g^2(x^j_{\alpha}-x^r_{\beta};  |n_r-n_j|+2k, v_r v_j),
\end{eqnarray}
where $M_r$ is the number of $r$-strings and
\begin{eqnarray}
\varphi_j(x)= \pi n_j(x+\frac{1-v_j}{2}i)+2n_j \phi i,  \quad
  g(x; n, v) = \frac{\sigma[\frac{i}{2}(x-n\eta_m i+ \frac{1-v}{2}i)]}{\sigma [\frac{i}{2}(x+n\eta_m i + \frac{1-v}{2}i)]}.\label{gdef}
\end{eqnarray}
Taking the logarithm of Eq.(\ref{BAE3}), we have
\begin{eqnarray}\label{count1}
  \frac{1}{i}\varphi_j(x^j_{\alpha})+N\vartheta_j (x^j_{\alpha})=2\pi I^j_{\alpha} + \sum^{z_l+1}_{r=1} \sum^{M_r}_{\beta=1}\Theta_{jr}(x^j_{\alpha}-x^r_{\beta}),
\end{eqnarray}
where $I^j_{\alpha}$ is an integer (or half-odd integer) for $ n_j+M_j+1-N\frac{1+v_j}{2} $ even (or odd) and
\begin{eqnarray}
&&  \vartheta_j (x) =\vartheta(x;  n_j, v_j)= -i \ln [(-v_j)g(x;  n_j, v_j)], \label{thetadef1}\\[4pt]
&&  \Theta_{jr}(x) =\vartheta(x;  n_r+n_j, v_jv_r) +\vartheta(x;  |n_r-n_j|, v_jv_r)\nonumber\\
  &&\qquad\qquad  +2 \sum^{\text{min}(n_r, n_j)-1}_{k=1}\vartheta(x;  |n_r-n_j|+2k, v_jv_r). \label{thetadef2}
\end{eqnarray}
The $\vartheta_j (x)$ is the elliptic function with double quasi-periodicities $2i$ and $2\frac{\tau}{i}$.
It is sufficient to consider their values in one periodicity. In order to shift the variables
into the region of $[-i, i]$, we define
\begin{eqnarray}
&&q_j \equiv (-1)^s (p_s -(j-z_s)p_{s+1}), \quad s=0, 1, \cdots, l, \quad j=1, 2, \cdots, z_l, \quad  z_s \leq j <z_{s+1},
 \nonumber\\
  && q_{z_l+1}=(-1)^{l+1} p_{l+1}, \label{pf}
\end{eqnarray}
where the series $\{p_s\}$ are
\begin{eqnarray}
&&p_0=\frac{N}{2m},\quad p_1=1, \quad p_n=p_{n-2} -p_{n-1}a_{n-1},  \nonumber  \\
&&a_{n-1}= \left\lfloor \frac{p_{n-2}}{p_{n-1}} \right\rfloor, \quad n=2, 3, \cdots, l+1. \label{p11f}
\end{eqnarray}
We note that the series $\{p_s\}$ are completely determined by the SCF expansion of $\frac{2m}{N}$ (\ref{scf}).
Thus $p_{l+1}=0$. From Eq.(\ref{pf}), we know $|q_r| \leq  |q_j|$ if $r>j$, and
$-1 \leq q_j \leq \frac{N}{2m}-1$ for $j=1, 2, \cdots, z_l +1$. These results can be seen clearly
from Table \ref{table}. In Appendix B, we prove that $q_j$
can also be parameterized as
\begin{eqnarray}
&& q_j = \omega_j \frac{N}{2m} -n_j, \nonumber \\
&& \omega_{z_1}=0, \quad \omega_j= \lfloor (n_j-1)\frac{2m}{N} \rfloor +1, \quad j\neq z_1. \label{qwr2}
\end{eqnarray}
We see that the quantity $q_j$ and integer $\omega_j$ are only determined by the length $n_j$ of $j$-string.
By using $q_j$ and integer $\omega_j$ in (\ref{qwr2}),
we express $\vartheta_j (x)$ as
\begin{eqnarray}\label{thetadef3}
 \vartheta_j (x) =\frac{1}{i}\ln \frac{\theta_{10}[ \frac{i}{2}(x+ ( q_j\frac{2m}{N}i +\frac{n_j}{N}\frac{\tau}{i} ))]}{\theta_{10}[\frac{i}{2}(x- ( q_j\frac{2m}{N}i +\frac{n_j}{N}\frac{\tau}{i} ))]},
\end{eqnarray}
where $\theta_{10}(x)$ is the elliptic theta function defined in Appendix A.
Then we conclude that the function $\vartheta_j(x)$ is a monotonically increasing function of $x$ for $q_j >0$ and a monotonically decreasing function of $x$ for $q_j<0$.

Substituting Eq.(\ref{str}) into the selection rule (\ref{BAE22}) and taking the logarithm, we have
\begin{eqnarray}\label{BAE2f}
  \sum^{z_l+1}_{j=1}  \left\{ \sum^{M_j}_{\alpha=1} \frac{1}{2\pi} \vartheta_j (x^j_{\alpha})  \right\} = I + \frac{k_1}{2N} -\frac{\phi}{2\pi}, \quad k_1 = 1,\cdots, 2N,
\end{eqnarray}
where $I$ is an integer (of half-odd integer) for $\sum^{z_l+1}_{j=1}  M_j(1+v_j)/2$ even (or odd).
From above equation, we see that the value of $\phi$ and the distributions of Bethe roots quantifying by function $\vartheta_j (x)$ are not independent.
For each given states, the distributions of strings are determined, thus the value of $\phi$ is determined.

Define the counting function $Z_j(x)$ as
\begin{eqnarray}
\label{count}
Z_j(x)=\frac{\varphi_j(x)}{2\pi N i}+\frac{1}{2\pi}\vartheta_j (x)- \frac{1}{N}\sum^{z_l+1}_{r=1} \sum^{M_r}_{\beta=1}\frac{1}{2\pi} \Theta_{jr}(x-x^r_{\beta}).
\end{eqnarray}
It is clear that $Z_j(x^j_{\alpha})=\frac{I^j_{\alpha}}{N}$ corresponds to the Eq.(\ref{count1}).
In the thermodynamic limit $N \rightarrow \infty$, the distribution of Bethe roots tends to continuous.
Thus $x^j_{\alpha}$ becomes a continuous variable and the counting function $Z_j(x)$ becomes a continuous function.
Define
\begin{eqnarray}
\frac{d } {dx} Z_j(x)= \text{sign}(q_j)[\rho_j(x)+\rho^h_j(x)],
\end{eqnarray}
where $\rho_j(x)$ is the density of states characterized by $j$-string, $\rho^h_j(x)$ is the density of corresponding holes, and
the sign function $\text{sign}(q_j)$ is added because of the monotonicity of function $\vartheta_j (x)$.

Taking the derivative of Eq.(\ref{count}) with respect to $x$,
we obtain the integral equations for the densities of states
\begin{eqnarray}\label{eq2}
  \text{sign}(q_j)\rho^h_j(x)=a_j(x) +\frac{n_j}{2Ni}-\sum^{z_l+1}_{r=1}\int^{Q}_{-Q}A_{jr}(x-y)\rho_r(y)dy,
\end{eqnarray}
where $Q$ is the integral bound and the functions $a_j(x)$, $A_{jr}(x)$ are
\begin{eqnarray}
&&  a_j(x) = \frac{1}{2\pi}\frac{d}{dx}\vartheta_j (x)=a(x; n_j, v_j) \nonumber\\[4pt]
&& \qquad\;\; = - \frac{1}{4\pi} \left\{  \frac{\theta'_{10}[\frac{i}{2}(x- ( q_j \frac{2m}{N}i +\frac{n_j\tau}{Ni} ))]}{\theta_{10}[\frac{i}{2}(x- (q_j \frac{2m}{N}i +\frac{n_j\tau}{Ni} ))]} - \frac{\theta'_{10}[\frac{i}{2}(x+ (q_j \frac{2m}{N}i +\frac{n_j\tau}{Ni} ))]}{\theta_{10}[\frac{i}{2}(x+ (q_j \frac{2m}{N}i +\frac{n_j\tau}{Ni} ))]}  \right\},\label{adef}\\[4pt]
&&   A_{jr}(x) = \frac{1}{2\pi}\frac{d}{dx}\Theta_j(x) + \delta_{jr}\text{sign}(q_j)\delta(x) \nonumber\\ [4pt]
&& \qquad\quad =a(x; n_r+n_j, v_rv_j)+ a(x; |n_r-n_j|,v_rv_j) \nonumber\\
  &&\qquad\qquad+2\sum_{k=1}a(x; |n_r-n_j|+2k,v_rv_j) + \delta_{jr}\text{sign}(q_j)\delta(x).\label{Adef}
\end{eqnarray}
The functions $a_j(x)$ and $A_{jr}(x)$ are elliptic functions with double-periodic $2i$ and $2\frac{\tau}{i}$.
Since the $j$-strings are distributed in the interval $[-\frac{\tau}{i},\frac{\tau}{i}]$ on the real axis (noting that $\tau$ is pure imaginary), we can choose $Q=\frac{\tau}{i}$.
%
%For a general bound state, the $j$-string are distributed in a finite region thus $Q\leq \frac{\tau}{i}$.
%In the following, we chose $Q$ as $\frac{\tau}{i}$, then the corresponding holes must be added and the density of holes must be introduced.
%The results obtained by these two methods are the same.

\subsection{Ground state energy}
\label{sec-ge}

Now, we are ready to calculate the physical quantities in the thermodynamic limit.
Since the real Bethe roots contribute negative energies, the Bethe roots should fill the real axis as far as possible at the ground state.
In general, the maximum number of real Bethe roots is $\frac{N}{2}$, and the remaining roots will be repelled to the complex plane and form the strings
satisfying the rule (\ref{str-rule}).

We express the real Bethe roots as $1$-strings, which correspond $n_j=1$ and $v_j=+1$ in Eq.(\ref{str-rule}).
From (\ref{Adef}), we obtain the function $A_{1r}(x)$ at the ground state
\begin{eqnarray}
  A_{1r}(x) = a(x; n_r+1, v_r) + a(x; n_r-1, v_r) + \delta_{1r}\text{sign}(q_1) \delta(x).
\end{eqnarray}
Above equation can be solved by the Fourier transformation. The Fourier transformation of a periodic function $F(x)$ is define as
\begin{eqnarray}
  \tilde{F}(k)= \int^{T}_{-T} F(x)e^{-ik\frac{\pi}{T}x} dx , \quad
  F(x) = \frac{1}{2T} \sum^{\infty}_{k=-\infty} \tilde{F}(k)e^{ik\frac{\pi}{T}x}, \quad k \in \textbf{Z}, \nonumber
\end{eqnarray}
where $x\in[-T, T]$ and $2T$ is the periodicity. The Fourier transformation of function $A_{1r}(x)$ is
\begin{eqnarray}
  \tilde{A}_{1r}(k) = 2\cosh(\frac{i}{\tau} k\pi \eta_m) \frac{\sinh [\frac{i}{\tau} k\pi (q_r\frac{2m}{N}- n_r\frac{\tau}{N})] }{\sinh(\frac{i}{\tau} k\pi)}.\label{tanq-1}
\end{eqnarray}
Taking the Fourier transformation of Eq.(\ref{eq2}) with $j=1$ and using Eq.(\ref{tanq-1}),
we obtain the densities of states at the ground state
\begin{eqnarray}\label{rhof}
  \tilde{\rho}_1(k) = \frac{1}{2\cosh(\frac{i}{\tau} k\pi \eta_m)} - \frac{\tilde{\rho}^h_1(k)}{\tilde{A}_{11}(k)}
  - \frac{ \tau \delta_{k0}}{N \tilde{A}_{11}(k) } -\sum^{z_l+1}_{r \neq 1}
  \frac{\sinh [\frac{i}{\tau} k\pi (q_r\frac{2m}{N}- n_r\frac{\tau}{N})] }{\sinh [\frac{i}{\tau} k\pi(1-\eta_m)]} \tilde{\rho}_r(k).
\end{eqnarray}
We see that the real Bethe roots, holes and strings are coupled together, thus
the distribution of real Bethe roots depends on the densities of holes and strings.

From Eqs.(\ref{eigE2}) and (\ref{adef}), the energy of $r$-string is
\begin{eqnarray}
&&  \varepsilon_r (\eta_m)=
 \frac{\sigma(\eta_m)}{\sigma'(0)} \left\{ \frac{\theta'_{10}[\frac{i}{2}(x- ( q_r \frac{2m}{N}i +\frac{n_r\tau}{Ni} ))]}{\theta_{10}[\frac{i}{2}(x- (q_r \frac{2m}{N}i +\frac{n_r\tau}{Ni} ))]}
 - \frac{\theta'_{10}[\frac{i}{2}(x+ (q_r \frac{2m}{N}i +\frac{n_r\tau}{Ni} ))]}{\theta_{10}[\frac{i}{2}(x+ (q_r \frac{2m}{N}i +\frac{n_r\tau}{Ni} ))]} \right\}  \nonumber\\[4pt]
&&\qquad = -4\pi \frac{\sigma(\eta_m)}{\sigma'(0)} a_r(x).\label{Estr}
\end{eqnarray}
From Eqs.(\ref{eigE2}), (\ref{rhof}) and (\ref{Estr}), the ground state energy reads
\begin{eqnarray}
&& E_g(\eta_m) =
-4\pi N \frac{\sigma(\eta_m)}{\sigma'(0)} \left\{ \int^{\frac{\tau}{i}}_{-\frac{\tau}{i}} a_1(x)\rho_1(x)dx \right.\nonumber\\
&&\quad\qquad\qquad \left. +  \sum^{z_l+1}_{r \neq 1} \int^{\frac{\tau}{i}}_{-\frac{\tau}{i}} a_r(x) \rho_r(x)dx \right\}
-i\pi \frac{\sigma(\eta_m)}{\sigma'(0)} +\frac{N}{2}\frac{\sigma'(\eta_m)}{\sigma'(0)} \nonumber\\
&&\qquad \quad \;\;= -\frac{2i \pi N}{\tau} \frac{\sigma(\eta_m)}{\sigma'(0)} \sum^{\infty}_{k=-\infty} \left\{ \frac{\tilde{a}_1(k)}{2\cosh(\frac{i}{\tau} k\pi\eta_m)} - \frac{ \tilde{\rho}^h_1(k)}{2\cosh(\frac{i}{\tau} k\pi\eta_m)} \right\}  +\frac{N}{2}\frac{\sigma'(\eta_m)}{\sigma'(0)}\nonumber\\[4pt]
  &&\qquad \quad\;\; = e_0(\eta_m) N +\epsilon_h(\eta_m) , \label{Eg}
\end{eqnarray}
where $e_0(\eta_m)$ is the density of ground state energy at degenerate point $\eta_m$ and $\epsilon_h(\eta_m)$ is the energy carried by
the holes in the real axis
\begin{eqnarray}
&&e_0(\eta_m) = -\frac{i \pi }{\tau} \frac{\sigma(\eta_m)}{\sigma'(0)} \sum^{\infty}_{k=-\infty} \frac{\sinh [\frac{i}{\tau} k\pi(1-\eta_m)]}{\sinh(\frac{i}{\tau} k\pi) \cosh(\frac{i}{\tau} k\pi \eta_m)}   +\frac{1}{2}\frac{\sigma'(\eta_m)}{\sigma'(0)} , \label{Eden}\\
&&  \epsilon_h(\eta_m)  = \frac{i \pi N }{\tau} \frac{\sigma(\eta_m)}{\sigma'(0)}\sum^{\infty}_{k=-\infty} \frac{\tilde{\rho}^h_1 (k)}{\cosh(\frac{i}{\tau} k\pi \eta_m)}.\label{Eh}
\end{eqnarray}
From Eq.(\ref{Eg}), we find that the ground state energy is only related to the real Bethe roots and corresponding holes.
Although the strings could affect the densities of sates,
their contribution to the energies is zero. This is because of the rearrangement of Fermi sea.
From Eqs.(\ref{Eden}) and (\ref{Eh}), we know that the ground state energy density is negative while the energy of hole is positive. Thus at the
ground state, the number of holes should be as less as possible to minimize the energy.
Due to the constraints of the BAEs, the density of holes should satisfy (\ref{rhof}).

As we mentioned before, the strings can affect the distribution of holes. Now, let us analyze the string solutions satisfying the rule (\ref{str-rule}).
Suppose at the ground state, there are $M_r$ $r$-strings $(r\geq 2)$.
From Eq.(\ref{pf}), we know $-1 \leq q_r \leq \frac{N}{2m}-1$ for $r=2, \cdots, z_l +1$.
Because we have shifted all the variables into one periodicity to compute the values of elliptic functions,
these strings should also be moved to the same periodicity. In order to minimize the energy, we require
\begin{eqnarray}\label{qjg}
  -1 < \sum^{z_l+1}_{r\neq 1} M_r q_r  < \frac{N}{2m}-1.
\end{eqnarray}
Based on Eq.(\ref{qjg}), it is straightforward that
\begin{eqnarray}
\left\lfloor (\sum^{z_l+1}_{r\neq 1} M_r n_r -1)\frac{2m}{N} \right\rfloor +1 = \sum^{z_l+1}_{r\neq 1}M_r\omega_r, \label{allw}
\end{eqnarray}
where $\sum^{z_l+1}_{r\neq 1} M_r n_r$ is the total number of string solutions. Substituting Eq.(\ref{allw}) into (\ref{qwr2}), we arrive at
\begin{eqnarray}
  \sum^{z_l+1}_{r\neq 1} M_r q_r =\left( \left\lfloor  (\sum^{z_l+1}_{r\neq 1} M_r n_r -1)\frac{2m}{N}\right\rfloor+1 \right) \frac{N}{2m} -\sum^{z_l+1}_{r\neq 1} M_r n_r . \label{strqr}
\end{eqnarray}
It is clear that the contribution of strings depends on the parity of system-size $N$.
Thus the number of holes and the ground state energy depend on the parity of $N$, and we should consider them separately.

If $N$ is odd. At the ground state, there are $M_1=\frac{N-1}{2}$ real Bethe roots and $\frac{N+1}{2}$ string solutions
\begin{eqnarray}\label{strlenoodd}
 \sum^{z_l+1}_{r\neq 1} M_r n_r = \frac{N+1}{2}.
\end{eqnarray}
Substituting (\ref{strlenoodd}) into (\ref{strqr}), we have
\begin{eqnarray}\label{strqrodd}
\sum^{z_l+1}_{r\neq 1} M_r q_r =  \left( \left\lfloor ( \frac{N+1}{2}-1)\frac{2m}{N} \right\rfloor+1 \right) \frac{N}{2m} -\frac{N+1}{2}=-\frac{1}{2}.
\end{eqnarray}
Because the number of real Bethe roots is $M_1=\frac{N-1}{2}$, we have
\begin{eqnarray}
\frac{M_1}{N} = \frac{1}{2}-\frac{1}{2N} =\int^{\frac{\tau}{i}}_{-\frac{\tau}{i}} \rho_1(x) dx = \tilde{\rho}_1(0). \label{dishole-90}
\end{eqnarray}
From the density of states (\ref{rhof}), we obtain the values of $\tilde{\rho}_1(0)$ as
\begin{eqnarray}
\tilde{\rho}_1(0)= \frac{1}{2} -\frac{\tilde{\rho}^{h}_1(0)}{2(1-\eta_m)} + \frac{ \eta_m}{2 N(1-\eta_m)}.\label{dishole-91}
\end{eqnarray}
Substituting (\ref{dishole-91}) into (\ref{dishole-90}), we obtain
\begin{eqnarray}\label{dishole2}
  \tilde{\rho}^h_1(0)= \int^{\frac{\tau}{i}}_{-\frac{\tau}{i}} \rho^h_1(x) dx = \frac{1}{N}.
\end{eqnarray}
Comparing this relation with the definition of counting function, we know that
such a configuration gives that there is only one hole in the real axis at the ground state.
The density of holes $\rho^{h}_1(x)$ can be expressed by the $\delta$-function as
\begin{eqnarray}\label{denh2}
  \rho^{h}_1(x) = \frac{1}{N} \delta(x-x^h), \quad  \tilde{\rho}^{h}_1(k)=\frac{1}{N}e^{\frac{k\pi}{\tau}x^h},
\end{eqnarray}
where $x^h$ is the position of hole.
Substituting (\ref{denh2}) into (\ref{Eh}), we obtain the energy carried by one hole
\begin{eqnarray}
 \epsilon_h(x^h,\eta_m) = \frac{i \pi }{\tau} \frac{\sigma(\eta_m)}{\sigma'(0)}\sum^{\infty}_{k=-\infty} \frac{ e^{k\pi x^h/\tau}}{\cosh(\frac{i}{\tau} k\pi \eta_m)}.\label{Eh3}
\end{eqnarray}
In the thermodynamic limit, the position $x^h$ of hole can take continuous values in the interval $[-\frac{\tau}{i} , \frac{\tau}{i}]$.
It should be noted that the function $\epsilon_h(x^h,\eta_m)$  takes the minimum value at $x^h=\frac{\tau}{i}$, which corresponding to the ground state.
Thus the ground state energy reads
\begin{eqnarray}
E^{odd}_g (\eta_m) =  e_0(\eta_m) N +\epsilon_h(\frac{\tau}{i},\eta_m). \label{Egodd}
\end{eqnarray}

If the system size $N$ is even, there are $M_1=\frac{N}{2}$ real Bethe roots and $\frac{N}{2}$ string solutions
at the ground state
\begin{eqnarray}\label{strleneven}
 \sum^{z_l+1}_{r\neq 1} M_r n_r = \frac{N}{2}.
\end{eqnarray}
Substituting (\ref{strleneven}) into (\ref{strqr}), we have
\begin{eqnarray}\label{strqreven}
\sum^{z_l+1}_{r\neq 1} M_r q_r =   \left(  \left\lfloor (\frac{N}{2}-1 )\frac{2m}{N} \right\rfloor+1 \right) \frac{N}{2m} -\frac{N}{2}=0.
\end{eqnarray}
From the Eqs.(\ref{rhof}), (\ref{strleneven}) and (\ref{strqreven}), we have
\begin{eqnarray}
  \frac{M_1}{N} = \frac{1}{2} =\int^{\frac{\tau}{i}}_{-\frac{\tau}{i}} \bar \rho_1(x) dx= \tilde{\bar \rho}_1(0)
= \frac{1}{2}-\frac{\tilde{\bar \rho}^h_1(0)}{2(1-\eta_m)},
\end{eqnarray}
where $\bar \rho_1(x)$ is the density of real Bethe roots and $\tilde{\bar \rho}_1(k)$ is the density of corresponding holes.
Such a configuration gives that there is no hole at the ground state, i.e., $\bar \rho^h_1(x)=0$.
Then the ground state energy is
\begin{eqnarray}\label{Egeven}
  E^{even}_g(\eta_m)= e_0(\eta_m) N.
\end{eqnarray}

\subsection{Elementary excitations}

Now, we consider a typical elementary excitations that is the hole excitation. The hole excitations of present model also have the parity. We first consider the odd $N$ case.
Obviously, the simplest excitation is putting one hole in the real axis, where the position of hole deviates from $\tau/i$.
The corresponding excited energy is quantified by
\begin{eqnarray}\label{Ecodd}
  \Delta E^{odd}(x^h, \eta_m)= E^{odd}_e(\eta_m)-E^{odd}_g(\eta_m)
  = \epsilon_h(x^h, \eta_m) -\epsilon_h(\frac{\tau}{i},\eta_m),
\end{eqnarray}
where $E_e^{odd}(\eta_m)$ is the energy at the excited state.
In the thermodynamic limit, the hole can move smoothly in the real axis which means
$x^h$ can tend to $\tau/i$ infinitely. Then we conclude that the excitation spectrum is continuous
\begin{eqnarray}\label{gapless}
\Delta E^{odd} (\eta_m) =  \lim_{x^h\rightarrow \tau/i } \Delta E^{odd} (x^h, \eta_m) \rightarrow 0.
\end{eqnarray}

The simplest hole excitation for even $N$ case is that a
real Bethe root is replaced by a $z_1$-string in the configuration of Bethe roots at the ground state.
The length of $z_1$-string is $n_{z_1}=1$ and the corresponding parity is $v_{z_1}=-1$, which is shown as the ninth column in Table \ref{table}
and the red dot in Fig.\ref{fig4}.
The energy carried by $z_1$-string is positive. In this kind of excited state,
there are $M_1=(\frac{N}{2}-1)$ real Bethe roots and
and $(\frac{N}{2}+1)$ string solutions.
The real Bethe roots satisfies
\begin{eqnarray}
\frac{M_1}{N}= \frac{1}{2}-\frac{1}{N}= \int^{\frac{\tau}{i}}_{-\frac{\tau}{i}} \rho'_1(x) dx=\tilde{\rho'}_1(0), \label{pf-90}
\end{eqnarray}
and the string solutions satisfy
\begin{eqnarray}
&& \sum^{z_l+1}_{r\neq 1} M_r n_r = \frac{N}{2}+1, \label{strleneve} \\
&& \sum^{z_l+1}_{r\neq 1} M_r q_r =    \left( \left\lfloor (\frac{N}{2}-1 )\frac{2m}{N} \right\rfloor+1 \right) \frac{N}{2m} -\frac{N}{2} + q_{z_1}=-1.\label{strqreve}
\end{eqnarray}
Due to the constraints (\ref{strleneve}) and (\ref{strqreve}), the density of states (\ref{rhof}) with $k=0$ reads
\begin{eqnarray}
\tilde{\rho'}_1(0)= \frac{1}{2}-\frac{\tilde{\rho'}^h_1(0)}{2(1-\eta_m)} + \frac{ \eta_m }{N(1-\eta_m)}. \label{pf-91}
\end{eqnarray}
Substituting (\ref{pf-91}) into (\ref{pf-90}), we obtain
\begin{eqnarray}
  \tilde{\rho'}^h_1(0)= \int^{\frac{\tau}{i}}_{-\frac{\tau}{i}} \rho'^h_1(x) dx = \frac{2}{N},
\end{eqnarray}
which indicates that
there exist two holes in the real axis. The density of holes can be expressed as
\begin{eqnarray}\label{denh}
  \rho'^h_1(x) = \frac{1}{N}[\delta(x-x^h_1) +\delta(x-x^h_2)],
\end{eqnarray}
where $x^h_1$ and $x^h_2$ are the positions of holes.
The energy of hole excitation is defined as
\begin{eqnarray}\label{Ecevenc}
\Delta E^{even}(x^h_1, x^h_2, \eta_m)=E_e^{even}(\eta_m) - E^{even}_g(\eta_m) =\epsilon_h(x^h_1,\eta_m)+\epsilon_h(x^h_2,\eta_m),
\end{eqnarray}
where $E_e^{even}(\eta_m)$ is the energy at excited state.
The values of $\Delta E^{even}(x^h_1, x^h_2, \eta_m)$ can be minimized by putting two holes at the point of $\tau/i$ in the thermodynamic limit.
Then we have
\begin{eqnarray}\label{gap}
\Delta E^{even}(\eta_m) = \lim_{x^h_1, x^h_2 \rightarrow \tau/i } \Delta E^{even}(x^h_1,x^h_2,\eta_m) =2\epsilon_h(\frac{\tau}{i},\eta_m),
\end{eqnarray}
which means that the hole excitation has a gap.

\section{Thermodynamic limit with arbitrary couplings}
\setcounter{equation}{0}
\label{sec-Phy}
%%%%%%%%%%%%%%% %%%%%%%  %%%%%%%%%%%%%  %%%%%%%%%%%%%%%%%%%

\subsection{Main ideas}

In this section, we generalize above results from degenerate points $\eta_m$ to the arbitrary real $\eta$ in the interval $0<Re(\eta)\leq \frac 12$.
The main idea is as follows. In principle, a physical quantity $E(\eta)$ which is a function of model parameter $\eta$ can be expressed as
\begin{eqnarray}
E(\eta)=N f_0(\eta) +f_1(\eta) + \frac{1}{N}f_2(\eta) +O(N^{-2}),\label{g22ap-1}
\end{eqnarray}
where $f_n(\eta)$ $ (n=0,1,2)$ are some unknown functions and $O(N^{-2})$ means the corrections up to the order of $N^{-2}$.
We do not know the explicit forms of $E(\eta)$ and $f_n(\eta)$ for the general $\eta$. What we know is the value of $E(\eta)$ at the point of $\eta_m$
\begin{eqnarray}
E(\eta_m)=N e_0(\eta_m) +e_1(\eta_m) + \frac{1}{N}e_2(\eta_m) +O(N^{-2}),\label{g22sfddfap-1}
\end{eqnarray}
where the functions $e_n(\eta_m)$ $ (n=0,1,2)$ have been determined. With the changing of $m$, there are a series of degenerate points $\{\eta_m\}$.
Substituting the $\eta=\eta_m, \eta_{m+1}$ into (\ref{g22ap-1}) and comparing with (\ref{g22sfddfap-1}), we obtain
\begin{eqnarray}
f_n(\eta_m)=e_n(\eta_m), \quad  f_n(\eta_{m+1})=e_n(\eta_{m+1}).
\end{eqnarray}
From Eq.(\ref{eta}), the difference between two
degenerate points $\eta_m$ and $\eta_{m+1}$ is proportional to $N^{-1}$, which means that the
$\eta_{m+1}-\eta_{m}$ is a small quantity with the order of $N^{-1}$.

Now, we consider a generic $\eta$ where $\eta_m \leq \eta \leq \eta_{m+1}$.
Suppose both $f_n(\eta)$ and $e_n(\eta)$ are the smooth functions of $\eta$.
We take the Taylor expansions of functions $f_n(\eta)$ and $e_n(\eta)$ at the point of $\eta_m$
\begin{eqnarray}
&&  f_n(\eta) = f_n(\eta_m) +f'_n(\eta_m)(\eta-\eta_m) + O(N^{-2}),\label{tayr3}\\[4pt]
&&  e_n(\eta) = e_n(\eta_m) +e'_n(\eta_m)(\eta-\eta_m) + O(N^{-2}).\label{tayr4}
\end{eqnarray}
Substituting $\eta=\eta_{m+1}$ into Eqs.(\ref{tayr3})-(\ref{tayr4}), we have
\begin{eqnarray}
&&  f_n(\eta_{m+1}) = f_n(\eta_{m}) +f'_n(\eta_{m})(\eta_{m+1}-\eta_{m}) + O(N^{-2}),\label{tayr2}\\[4pt]
&&  e_n(\eta_{m+1}) = e_n(\eta_{m}) +e'_n(\eta_{m})(\eta_{m+1}-\eta_{m}) + O(N^{-2}). \label{tayr1}
\end{eqnarray}
Eq.(\ref{tayr2}) minus (\ref{tayr1}) gives
\begin{eqnarray}
f'_n(\eta_{m}) = e'_n(\eta_{m}) + O(N^{-1}).\label{tayr81}
\end{eqnarray}
Substituting (\ref{tayr81}) into (\ref{tayr3}), we have
\begin{eqnarray}
f_n(\eta) =  e_n(\eta_m) +e'_n(\eta_m)(\eta-\eta_m)+ O(N^{-2})
=e_n(\eta) + O(N^{-2}),
\end{eqnarray}
which means that the functions $f_n(\eta)$ can be obtained from the determined ones $e_n(\eta)$ up to the order of $N^{-2}$.
In the thermodynamic limit, the correction $O(N^{-2})$ can be neglected and the
physical quantity $E(\eta)$ can be obtained from $E(\eta_m)$ with great accuracy.

Some remarks are in order. Because the degenerate point $\eta_m$ contains an imaginary part $\tau/N$,
the above extrapolations are valid for the line $\tau/N$ in the complex plain with varying the value of $m$. That is to say, we obtain the physical quantities along this line.
In the thermodynamic limit, we have $\tau/N \rightarrow 0$ thus can study the physics with real model parameter $\eta$, i.e.,  $0< \eta \leq \frac{1}{2}$
due to $0< \frac{2m}{N} \leq \frac{1}{2}$. However, our method also allows us to study the thermodynamic limit of the XYZ model with $\eta$ taking others values beyond a real one.
If $l_1$ tends to infinite while $l_1/N$ is finite in the thermodynamic limit,
the degenerate points are $\eta_{m, l_1}= -\frac{2l_1+1}{N}\tau +\frac{2m}{N}$, where
we have put $M=N$ and $m_1=-m$ in Eq.(3.1). Because $\tau$ is pure imaginary, the $\eta_{m, l_1}$ is complex.
We note that the above method can also be applied directly and one can obtain the corresponding results for the generic
complex $\eta$ with a finite imaginary part.
If $\eta$ is pure imaginary, we can put $M=N$ and $m_1=0$ in Eq.(3.1). Then with the changing of $l_1$, there exist sufficient degenerate points in the thermodynamic limit.
Based on them, we can obtain the associated results similarly.

%
%
%If we consider the case that $l_1$ in (\ref{eta}) is fixed meanwhile $l_1/N$ is finite, then we can obtain the corresponding results for the generic complex $\eta$
%with finite imaginary part. If $\eta$ is pure imaginary, we can put $m_1=0$ in Eq.(\ref{eta}) and study the changing of $l_1$.
%The associated results can be obtained similarly.

\subsection{Ground state energy}

Using above technique, we obtain the ground state energy of Hamiltonian (\ref{Hamiltoniant}) with odd $N$ and generic $\eta$ as
\begin{eqnarray}\label{Egoddr}
E^{odd}_g(\eta)=e_0(\eta) N +\epsilon_h(\frac{\tau}{i},\eta),
\end{eqnarray}
where the density $e_0(\eta)$ and function $\epsilon_h(x,\eta)$ are given by (\ref{Eden}) and (\ref{Eh3}) provided that $\eta_m$ is replaced by $\eta$, respectively.

Next, we check the result (\ref{Egoddr}) by numerical calculations.
In the verifications, we randomly chose the values of model parameters $\tau$ and $\eta$. Meanwhile,
we require that $\eta\neq \eta_m$, because we have proved analytically that Eq.(\ref{Egoddr}) is correct if $\eta=\eta_m$.
In order to quantify the validity of Eq.(\ref{Egoddr}), we define
\begin{eqnarray}\label{dEg-oddt}
\delta^{odd} =  \frac{E^{odd}_{g}(\eta) }{\bar{E}^{odd}_{g}(\eta)}-1,
\end{eqnarray}
where $\bar{E}^{odd}_{g}(\eta)$ is the ground state energy of
Hamiltonian (\ref{Hamiltoniant}) obtained by the exact diagonalization and $ E^{odd}_{g}(\eta)$ is the ground state energy obtained by Eq.(\ref{Egoddr}).

In Table \ref{table-oddt}, we list the results with $\tau=0.5i$ and $\eta=0.4$. We see that the value of $\delta^{odd}$
has the order of $10^{-9}$ when the system size $N=25$, which indicates that Eq.(\ref{Egoddr}) can describe the ground state energy
with satisfied accuracy even for the system with small size.
\begin{table}[!htp]
  \caption{
Numerical results for the odd $N$. Here
$\bar{E}^{odd}_g$ is the ground state energy of Hamiltonian (\ref{Hamiltoniant}) obtained by the exact diagonalization with the help of the sparse matrix method,
$E^{odd}_g$ is the ground state energy obtained by Eq.(\ref{Egoddr}), $\delta^{odd}$ characterizes the difference, $\tau=0.5i$ and $\eta=0.4$. }
  \centering
  \begin{tabular}{| r|r|r|r|}\hline
$ N$ & $  \bar{E}^{odd}_{g}$ & $  E^{odd}_{g}$ & $ \delta^{odd}~(\times 10^{-3})$  \\ \hline
 $ 5 $ & $-4.24631809$  & $-4.25772417$  & $2.68611098$   \\ \hline
 $ 7 $ & $-6.58795792$  & $-6.59092288$  & $0.45005755$   \\ \hline
 $ 9 $ & $-8.92330609$  & $-8.92412158$  & $0.09138932$   \\ \hline
 $ 11 $ & $-11.25708902$  & $-11.25732029$  & $0.02054430$   \\ \hline
 $ 13 $ & $-13.59045214$  & $-13.59051899$  & $0.00491948$   \\ \hline
 $ 15 $ & $-15.92369811$  & $-15.92371770$  & $0.00122998$   \\ \hline
 $ 17 $ & $-18.25691061$  & $-18.25691640$  & $0.00031735$   \\ \hline
 $ 19 $ & $-20.59011338$  & $-20.59011511$  & $0.00008386$   \\ \hline
 $ 21 $ & $-22.92331330$  & $-22.92331381$  & $0.00002258$   \\ \hline
 $ 23 $ & $-25.25651236$  & $-25.25651252$  & $0.00000617$   \\ \hline
 $ 25 $ & $-27.58971118$  & $-27.58971122$  & $0.00000171$   \\ \hline
 \end{tabular}
\label{table-oddt}
\end{table}

\begin{figure}[!htp]
    \centering
    \includegraphics[height=4.5cm]{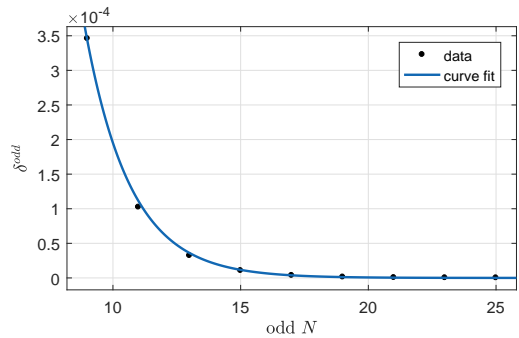}
    \caption{
The deviation $\delta^{odd}$ versus the system-size $N$.
The data can be fitted as $ \delta^{odd}=\alpha_1 \exp (\beta_1 N)$.
Here $\tau =0.5i$ $\eta=\frac{\sqrt{2}}{4}$, $\alpha_1=0.05447$ and $\beta_1=-0.5632$.
Due to the fact $\beta_1< 0$, the $\delta^{odd}$ tends to zero when $N\rightarrow \infty$.}\label{Eg-oddt}
\end{figure}

In Fig.\ref{Eg-oddt}, the finite size effect of $\delta^{odd}$ with $\tau=0.5i$ and $\eta= \sqrt{2}/4$ is given.
From the fitting, we find that $\delta^{odd}$ satisfies the exponential law
\begin{eqnarray}
\delta^{odd}=\alpha_1 \exp(\beta_1 N),
\end{eqnarray}
where $\alpha_1$ and $\beta_1$ are the fitting factors. Due to the fact $\beta_1< 0$,
the $\delta^{odd}$ tends to zero when the system size $N \rightarrow \infty$,
which indicates that Eq.(\ref{Egoddr}) gives the correct ground state energy
of antiperiodic XYZ spin chain in the thermodynamic limit.

The ground state energy of model (\ref{Hamiltoniant}) with even $N$ and generic $\eta$ is
\begin{eqnarray}\label{Egevenr}
 E^{even}_g(\eta) = e_0(\eta) N.
\end{eqnarray}
In order to check the validity of Eq.(\ref{Egevenr}), we define
\begin{eqnarray}\label{dEg-event}
\delta^{even} = \frac{E^{even}_{g}(\eta)}{\bar{E}^{even}_{g}(\eta)} -1,
\end{eqnarray}
where $\bar{E}^{even}_{g}(\eta)$ is the ground state energy obtained by the exact diagonalization
and $E^{even}_g(\eta)$ is the ground state energy obtained by Eq.(\ref{Egevenr}).

In Table \ref{table-event}, we list the numerical results with $\tau=0.5i$ and $\eta =0.4$.
The data show that the value of $\delta^{even}$ has the order of $10^{-15}$ when the system size $N=24$,
which indicates that Eq.(\ref{Egevenr}) can quantify the ground state energy
with satisfied accuracy even for the system with small size.
\begin{table}[!htp]
 \caption{
Numerical results for the even $N$. Here $\bar{E}^{even}_g$ is the ground state energy
obtained by the exact diagonalization,
$E^{even}_g$ is the ground state energy obtained by Eq.(\ref{Egevenr}),
$\delta ^{even}$ characterizes the difference, $\tau=0.5i$ and $\eta=0.4$.}
  \centering
\begin{tabular}{| r|r|r|r|}\hline
$ N$ & $ \bar{E}^{even}_{g} $ & $ E^{even}_{g}$ & $ \delta^{even}~(\times 10^{-4})$  \\ \hline
 $ 4 $ & $-4.66414993812$  & $-4.66639740988$  & $4.81860959229$   \\ \hline
 $ 6 $ & $-6.99945302886$  & $-6.99959611482$  & $0.20442450017$   \\ \hline
 $ 8 $ & $-9.33278427176$  & $-9.33279481977$  & $0.01130210061$   \\ \hline
 $ 10 $ & $-11.66599268867$  & $-11.66599352471$  & $0.00071664895$   \\ \hline
 $ 12 $ & $-13.99919216041$  & $-13.99919222965$  & $0.00004945726$   \\ \hline
 $ 14 $ & $-16.33239092868$  & $-16.33239093459$  & $0.00000361594$   \\ \hline
 $ 16 $ & $-18.66558963902$  & $-18.66558963953$  & $0.00000027571$   \\ \hline
 $ 18 $ & $-20.99878834443$  & $-20.99878834447$  & $0.00000002169$   \\ \hline
 $ 20 $ & $-23.33198704941$  & $-23.33198704941$  & $0.00000000174$   \\ \hline
 $ 22 $ & $-25.66518575436$  & $-25.66518575436$  & $0.00000000015$   \\ \hline
 $ 24 $ & $-27.99838445930$  & $-27.99838445930$  & $0.00000000002$   \\ \hline
 \end{tabular}
\label{table-event}
 \end{table}

\begin{figure}[!htp]
    \centering
    \includegraphics[height=4.5cm]{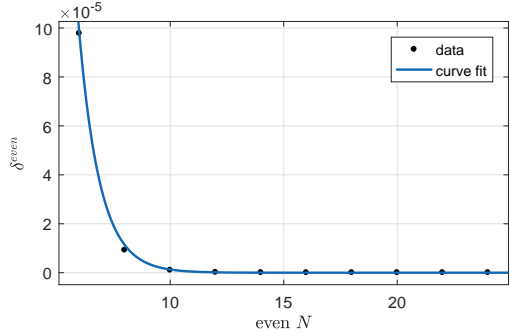}
    \caption{
The deviation $\delta^{even}$ versus the system-size $N$.
The data can be fitted as $ \delta^{even}=\alpha_2 \exp (\beta_2 N)$.
Here $\tau =0.5i$, $\eta=\frac{\sqrt{2}}{4}$, $\alpha_2=0.06319$ and $\beta_2 =-1.079$.
Due to the fact $\beta_2< 0$, the $\delta^{even}$ tends to zero when $N\rightarrow \infty$.}\label{Eg-event}
\end{figure}

In Fig.\ref{Eg-event}, we list the finite size effect of $\delta^{even}$ with $\tau=0.5i$
and $\eta= \sqrt{2}/4$. The data can be fitted as
\begin{eqnarray}
  \delta^{even}=\alpha_2 \exp (\beta_2 N).
\end{eqnarray}
Due to the fact $\beta_2< 0$, the $\delta^{even}$ tends to zero when the system size $N \rightarrow \infty$,
which indicates that Eq.(\ref{Egevenr}) gives the correct ground state energy
in the thermodynamic limit.

\subsection{Elementary excitations}

The energy of hole excitation with odd $N$ and generic $\eta$ is
\begin{eqnarray}\label{Ecoddr}
  \Delta E^{odd}(x^h, \eta) = \epsilon_h(x^h, \eta) -\epsilon_h(\frac{\tau}{i},\eta).
\end{eqnarray}
In the thermodynamic limit, the position of hole $x^h$ can take continuous values in the interval $[ -\frac{\tau}{i} , \frac{\tau}{i} ]$.
Then we have
\begin{eqnarray}\label{Efodd}
\Delta E^{odd} (\eta) =  \lim_{x^h\rightarrow \tau/i } \Delta E^{odd} (x^h, \eta) \rightarrow 0,
\end{eqnarray}
which indicates that the hole excitations are continuous.

\begin{figure}[!htp]
    \centering
    \includegraphics[height=5cm]{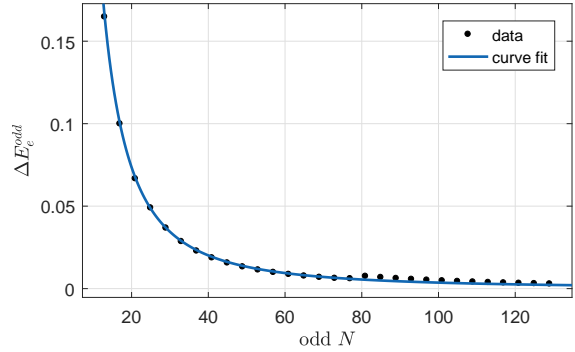}
    \caption{ The energy of hole excitation $\Delta E^{odd}_e$ versus the system-size $N$.
Here $\tau =0.5i$, $\eta=\frac{\sqrt{2}}{4}$, $\alpha_3 =19.87$ and $\beta_3 =-1.869$.
The data can be fitted as $ \Delta E^{even}_e=\alpha_3 N^{\beta_3}$.
Because of $\beta_3 < 0$, the $\Delta E^{odd}_e$ tends to zero when $N \rightarrow \infty$.}\label{exc-odd}
\end{figure}

Now, we check the result (\ref{Efodd}) by the density matrix renormalization group (DMRG) method \cite{white1993density, schollwock2005density}.
We use the infinite chain DMRG algorithm and start with 14 sites, where
the number of reserved states $m=2^7$ and the truncation error is $10^{-8}$.
We note that even in the small $N$ (less than 24) case, the DMRG results are in pretty good agreement with the direct diagonalization results,
where the relative errors at ground state is $10^{-9}$.
The numerical results, denoted as $\Delta E^{odd}_e(\eta)$, with $\tau=0.5i$ and $\eta=\sqrt{2}/4$ for various system size $N$
are shown in Fig.\ref{exc-odd}.
The data can be fitted as
\begin{eqnarray}\label{arodd}
  \Delta E^{odd}_e(\eta) =\alpha_3 N^{\beta_3}.
\end{eqnarray}
Due to the fact $\beta_3< 0$, the value $\Delta E^{odd}_e(\eta)$ tends to zero when the system size $N \rightarrow \infty$,
which is consistent with the analytical result (\ref{Efodd}).
%%%%%%%%%%%%%%%%%%%%%%%%%%%%%%%%%%%%%%%%%%

\begin{figure}[!htp]
    \centering
    \includegraphics[height=4.5cm]{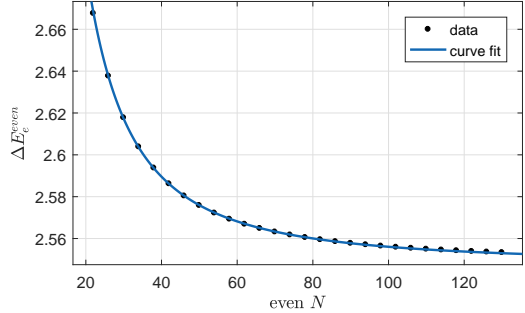}
    \caption{
The energy of hole excitation $\Delta E^{even}_e$ versus the system-size $N$.
Here $\tau =0.5i$, $\eta=\frac{\sqrt{2}}{4}$, $\alpha_4 =28.46$, $\beta_4 =-1.769$ and $\varepsilon_4 =2.548$.
The data can be fitted as $ \Delta E^{even}_e=\alpha_4 N^{\beta_4} +\varepsilon_4$.
Due to the fact $\beta_4 < 0$, $\epsilon_4$ gives the value of $\Delta E^{even}_e$ in the thermodynamic limit.}\label{exc-even}
\end{figure}

The energy of hole excitation with even $N$ and generic $\eta$ is
\begin{eqnarray}\label{Ecevenr}
\Delta E^{even}(x^h_1, x^h_2, \eta)= \epsilon_h(x^h_1, \eta)+\epsilon_h(x^h_2, \eta).
\end{eqnarray}
In the thermodynamic limit, two holes $x^h_1$ and $x^h_2$ can be put on the point of $\tau/i$ and we have
\begin{eqnarray}\label{Efeven}
\Delta E^{even}(\eta) = \lim_{x^h_1,x^h_2\rightarrow \tau/i } \Delta E^{even}(x^h_1,x^h_2,\eta) =2\epsilon_h(\frac{\tau}{i},\eta).
\end{eqnarray}
Now, we check the validity of Eq.(\ref{Efeven}).
Substituting $\tau=0.5i$ and $\eta=\frac{\sqrt{2}}{4}$ into Eq.(\ref{Efeven}) and using the definition of elliptic functions, we obtain
\begin{eqnarray}
\Delta E^{even}(\frac{\sqrt{2}}{4})= 2.54881. \label{areven}
\end{eqnarray}
We also check the analytical result (\ref{areven}) by DMRG methods.
The DMRG data, denoted as $\Delta E^{even}_e(\eta)$, with $\tau=0.5i$ and $\eta=\frac{\sqrt{2}}{4}$ for various even system size are shown in Fig.\ref{exc-even}.
The data can be fitted as
\begin{eqnarray}
\Delta E^{even}_e(\eta)=\alpha_4 N^{\beta_4} +\varepsilon_4. \label{aresdven}
\end{eqnarray}
Because of $\beta_4 < 0$, in the thermodynamic limit, $\varepsilon_4$ gives the values of
energy, i.e.,
\begin{eqnarray}
\varepsilon_4=\Delta E^{even}(\eta).
\end{eqnarray}
Meanwhile, the numerical calculation gives $\varepsilon_4 = 2.548$, which is consistent with the analytical result (\ref{areven}).

So far, we have obtained the ground state energy and hole excitation.
All these physical quantities are related to the parity of system size.
From Eqs.(\ref{Egoddr}) and (\ref{Egevenr}), we find that the ground state energy with odd $N$
includes a term $\epsilon_h(\frac{\tau}{i},\eta)$ induced by the hole in the real axis.
The magnitude of $\epsilon_h(\frac{\tau}{i},\eta)$ is the same as the ground state energy density $e_0(\eta)$.
From Eqs.(\ref{Efodd}) and (\ref{Efeven}), we find that the model with odd $N$ has a continuous hole excitation spectrum,
while the excitation for the even $N$ case has an energy gap $2\epsilon_h(\frac{\tau}{i},\tau)$.

\section{Limit to the antiperiodic XXZ model }
\setcounter{equation}{0}

A very strong motivation of this paper is to study the thermodynamic limit of the antiperiodic XXZ model.
Because the corresponding inhomogeneous BAEs do not have the degenerate points, the above method does not work.
In this section we show that the thermodynamic limit of the antiperiodic XXZ model can be obtained from the antiperiodic XYZ model.
The Hamiltonian of XXZ model reads
\begin{eqnarray}\label{Hxxz}
\bar H=\frac{1}{2} \sum^N_{j=1} \left[  \sigma^x_j \sigma^x_{j+1} +  \sigma^y_j \sigma^y_{j+1} + \cos(\pi \eta) \sigma^z_j \sigma^z_{j+1} \right],
\end{eqnarray}
and the antiperiodic boundary condition is given by Eq.(\ref{antic}).
The eigenvalue of Hamiltonian (\ref{Hxxz}) is\cite{cao2013off}
\begin{eqnarray}\label{Exxz}
\bar  E=- i\sin (\pi \eta) \sum^N_{j=1} [\coth(\lambda_j +i \pi \eta) -\coth (\lambda_j)] + \frac{N}{2} \cos (\pi \eta) + i \sin (\pi \eta),
\end{eqnarray}
where the $N$ Bethe roots $\{\lambda_j\}$ satisfy the BAEs
\begin{eqnarray}
  &&   e^{2\lambda_j+i \pi \eta} \frac{\sinh^N(\lambda_j +i \pi \eta)}{\sinh^N(\lambda_j )}
  = \prod^N_{k=1} \frac{\sinh(\lambda_j -\lambda_k +i\pi \eta)}{\sinh(\lambda_j -\lambda_k -i\pi \eta)} \nonumber \\
  && \qquad
  + c(\lambda_j)e^{\lambda_j+i \pi \eta}  \prod^N_{k=1} \frac{\sinh(\lambda_j + i \pi \eta)}{\sinh(\lambda_j -\lambda_k - i\pi \eta)}, \quad j=1, \cdots, N,  \label{BAExxz}
\end{eqnarray}
and $c(\lambda_j)$ is determined as
\begin{eqnarray}\label{cdef}
 c(\lambda_j)=e^{\lambda_j- i N\pi \eta -\sum^N_{l=1}\lambda_l } -e^{-\lambda_j -i \pi \eta +\sum^N_{l=1} \lambda_l }.
\end{eqnarray}
We see that the parameter $c(\lambda_j)$ and Bethe roos $\{\lambda_j\}$ are determined together and BAEs (\ref{BAExxz}) can not be reduced to the homogeneous ones.

It is well-known that the XXZ model can be obtained from XYZ model by taking the trigonometric limit
\begin{eqnarray}
\lim_{\tau \rightarrow i \infty} J_x(\eta, \tau)   \rightarrow 1,\quad
 \lim_{\tau \rightarrow i \infty} J_y(\eta, \tau)  \rightarrow 1,\quad \lim_{\tau \rightarrow i \infty} J_z(\eta, \tau) \rightarrow \cos(\pi \eta).
\end{eqnarray}
Therefore, some physics including the exact solution, BAEs, ground state, elementary excitation and thermodynamics of XXZ model
can be obtained from XYZ model by taking the same limit, provided that the limit exists.

In the following, we show that the thermodynamic limits of antiperiodic XXZ model indeed can be obtained from XYZ model.
Due to the constraint $0<\eta\leq \frac{1}{2}$, the related results are valid in the massless region $0<\cos(\pi \eta)\leq 1$ of XXZ model.
The ground state energy of antiperiodic XXZ model with odd $N$ is obtained by taking the limit $\tau \rightarrow i \infty$ of Eq.(\ref{Egoddr})
\begin{eqnarray}\label{Egxxz1}
  \bar E_g^{odd}(\eta)=\lim_{\tau \rightarrow i \infty }E^{odd}_g(\eta)  = \bar e_0( \eta) N +\bar \epsilon_h(\infty, \eta).
\end{eqnarray}
Here $\bar e_0( \eta)$ is the ground state energy density of the antiperiodic XXZ model
\begin{eqnarray}
 \bar e_0( \eta) = \lim_{\tau \rightarrow i \infty } e_0(\eta)
 = -\frac{ \sin(\pi \eta) }{\pi}  \int^{\infty}_{-\infty} \frac{\sinh [w(1-\eta)]}{\sinh(w) \cosh(w\eta)} d w  +\frac{1}{2}\cos(\pi \eta). \label{Edenxxz}
\end{eqnarray}
It is shown that  $\bar e_0( \eta)$ (\ref{Edenxxz}) obtained here is the same as that of the periodic XXZ model obtained in \cite{takahashi2005thermodynamics}.
The $\bar \epsilon_h(x^h, \eta)$ is the energy carried by one hole
\begin{eqnarray}
\bar \epsilon_h(x^h, \eta) = \lim_{\tau \rightarrow i \infty } \epsilon_h(x^h,\eta)
   =  \frac{\sin(\pi \eta)}{\eta}  \frac{ 1}{\cosh(\frac{\pi x^h}{2\eta})}, \label{Ehxxz}
\end{eqnarray}
where $x^h$ is the position of hole which is distributed in the interval $(-\infty, \infty)$.
If the hole is put on the infinity, from Eq.(\ref{Ehxxz}), we obtain that the minimum energy carried by hole is zero, i.e., $\bar \epsilon_h(\infty, \eta) =0$.

The ground state energy with even $N$ is obtained by taking the limit $\tau \rightarrow i \infty$ of Eq.(\ref{Egevenr}),
\begin{eqnarray}\label{Egxxz}
\bar E_g^{even}(\eta)=\lim_{\tau \rightarrow i \infty }E^{even}_g(\eta)  = \bar e_0( \eta) N.
\end{eqnarray}
From Eqs.(\ref{Egxxz1}) and (\ref{Egxxz}), we know that the ground state energy of antiperiodic XXZ model does not depend
on the parity of system size $N$ in the thermodynamic limit. That is to say, although the distributions of Bethe roots (or the densities of states) with odd and even $N$ are different,
the ground state energy can be expressed as a unified form of
\begin{eqnarray}
\bar E_g= \bar e_0( \eta) N.
\end{eqnarray}
This is because we consider the massless region and
the contribution of hole is zero.

The energy of hole excitation with odd $N$ is obtained by taking the limit $\tau \rightarrow i \infty$ of Eq.(\ref{Efodd}). Obviously, the hole excitation is continuous
\begin{eqnarray}\label{Exxzodd}
  \Delta \bar E^{odd}(\eta) \rightarrow 0.
\end{eqnarray}
From Eq.(\ref{Efeven}), we obtain the energy of hole excitation with even $N$
\begin{eqnarray}\label{Exxzeven}
\Delta \bar E^{even}(\eta)=\lim_{\tau \rightarrow i \infty } 2\epsilon_h(\frac{\tau}{i},\eta)  \rightarrow 0.
\end{eqnarray}
Therefore, the energy gap in the XYZ model
closes after taking the trigonometric limit.
Comparing Eqs.(\ref{Exxzodd}) and (\ref{Exxzeven}), we find that the hole excitation is gapless and the excited spectrum is continuous, which is true for both odd and even $N$.
We note that the physical pictures of excitations with odd and even $N$ are quite different, which have been explained in previous sections.

It is remarked that the process of taking limits in this paper is as follows. We first set $\tau$ to be finite, then we set $N$ to be infinite. The resulting degenerate points (\ref{eta2}) tend to real numbers which are dense in the real line. By this way, we obtain the thermodynamic limit of the XYZ model with a real $\eta$ and a finite $\tau$. The detailed results are given in sections 3 and 4. Then taking  the trigonometric limit of $\tau \rightarrow i\infty$ (or $\lim_{\tau\rightarrow  i\infty}\lim_{N\rightarrow \infty}\,\frac{\tau}{N}=0$ ), we obtain the corresponding results of the XXZ model with a real $\eta$ from those of the corresponding XYZ model.

\section{Conclusions}

We study the thermodynamic limit of the anisotropic spin-$\frac{1}{2}$ XYZ spin chain with the antiperiodic boundary condition described by the Hamiltonian (\ref{Ham-XYZ})
and (\ref{Antiperiodic}) based on its off-diagonal Bethe ansatz solution.
We overcome the difficult that it is hard to take the thermodynamic limit of the associated BAEs deriving from its inhomogeneous $T-Q$ relation.
With the help of  the exact results of system at degenerate points (\ref{eta2}), we obtain the
actual values of physical quantities such as the ground state energies and hole excitations for an arbitrary coupling $\eta$.
In this paper, we consider the case that
the anisotropic coupling parameter $\eta$ after taking thermodynamic limit is real one and $0<\eta\leq \frac{1}{2}$.
The results with generic complex model parameters including the pure imaginary ones can be derived similarly.
We also propose a method to study the thermodynamic limit of integrable model without degenerate points. As an example,
the results of the antiperiodic XXZ spin chain are obtained with the help of those of the antiperiodic XYZ spin chain.

The method proposed in this paper can be generalized to study the high rank quantum integrable models such as the $su(n)$ XYZ spin chain (or the $Z_n$-Belavin model) with antiperiodic
boundary conditions or with off-diagonal boundary reflections. In the thermodynamic limit, there are sufficient degenerate points where the
inhomogeneous Bethe ansatz equations reduce to the homogeneous ones. Based on them, we can study the ground state, elementary excitations and thermodynamic quantities.
Furthermore, the thermodynamic results of antiperiodic $su(n)$ XYZ model can give the corresponding results of the antiperiodic $su(n)$ XXZ model which does not have degenerate points.
We also expect that this method can be applied to other quantum integrable models.

\section*{Acknowledgments}

The financial supports from National Program for Basic Research of
MOST (Grant Nos. 2016 YFA0300600 and 2016YFA0302104), National
Natural Science Foundation of China (Grant Nos. 11934015,
11975183, 11947301, 11774397, 11775178 and 11775177),
Major Basic Research Program of Natural Science of Shaanxi
Province (Grant Nos. 2017KCT-12, 2017ZDJC-32), Australian Research
Council (Grant No. DP 190101529), Strategic Priority Research
Program of the Chinese Academy of Sciences (Grant No. XDB33000000) and Double First-Class
University Construction Project of Northwest University are
gratefully acknowledged.

\section*{Appendix A: Some elliptic theta functions}
\label{appA}
\setcounter{equation}{0}
\renewcommand{\theequation}{A.\arabic{equation}}

The elliptic $\theta$-function is defined by
\begin{eqnarray}\label{thef}
  \theta \left[
           \begin{array}{c}
             a \\
             b \\
           \end{array}
         \right](u,\tau)=\sum_m e^{\pi i(m+a)^2\tau +2\pi i(m+a)(u+b)},
\end{eqnarray}
where $a$ and $b$ are rational numbers, and $\tau$ is a generic complex number with $ \textrm{Im}(\tau)>0$.
For convenience, we adopt the following notations
\begin{eqnarray}
   && \theta_{11}(u)=\theta \left[
           \begin{array}{c}
             \frac{1}{2} \\
             \frac{1}{2} \\
           \end{array}
         \right](u,\tau),\quad
   \theta_{10}(u)=\theta \left[
           \begin{array}{c}
             \frac{1}{2} \\
             0 \\
           \end{array}
         \right](u,\tau),  \nonumber\\
    &&\theta_{00}(u)=\theta \left[
           \begin{array}{c}
             0 \\
             0 \\
           \end{array}
         \right](u,\tau),\quad
    \theta_{01}(u)=\theta \left[
           \begin{array}{c}
             0 \\
             \frac{1}{2} \\
           \end{array}
         \right](u,\tau). \label{thef2}
\end{eqnarray}
From the definition (\ref{thef}), we know that $\theta_{11}(u)$ is an odd function of $u$ and $\theta_{10}(u)$, $\theta_{00}(u)$, $\theta_{01}(u)$ are the even functions of $u$.
These functions are doubly quasi-periodic and satisfy
\begin{eqnarray}
&&  \theta_{11}(u+1) = -\theta_{11}(u), \quad \theta_{11}(u+\tau ) = - e^{-2i\pi (u+\frac{\tau}{2})} \theta_{11}(u),  \nonumber\\
&&  \theta_{10}(u+1) = -\theta_{10}(u), \quad  \theta_{10}(u+\tau) = e^{-2i\pi (u+\frac{\tau}{2})} \theta_{10}(u),  \nonumber\\
&&  \theta_{00}(u+1 )= \theta_{00}(u), ~\quad  \theta_{00}(u+\tau) = e^{-2i\pi (u+\frac{\tau}{2})} \theta_{00}(u),  \nonumber\\
&& \theta_{01}(u+1) = \theta_{01}(u), ~\quad   \theta_{01}(u+\tau) = - e^{-2i\pi (u+\frac{\tau}{2})} \theta_{01}(u).\label{theta-per}
\end{eqnarray}

\section*{Appendix B: Proof of relation (\ref{qwr2})}
\setcounter{equation}{0}
\renewcommand{\theequation}{B.\arabic{equation}}

It is convenient to introduce a series $\{y'_{-1}, y'_{0},  \cdots, y'_{l}\}$ as
\begin{equation}\label{yip}
  y'_{-1}=1, \quad y'_0=0, \quad y'_k=y'_{k-2}+ a_ky'_{k-1}, \quad k=1, 2, \cdots, l,
\end{equation}
where the integers $a_k$ are given by Eq.(\ref{scf}). Combining above definition and Eqs.(\ref{yf}) and (\ref{p11f}), we find
\begin{eqnarray}\label{pf2}
  (-1)^t p_t =y'_{t-1} p_0 -y_{t-1}, \quad t=0, 1, \cdots, l+1, \label{pf3}
\end{eqnarray}
Thus the $q_j$ given in (\ref{pf}) reads
\begin{eqnarray}
  q_j = \omega_j p_0 -n_j , \label{qjf3}
\end{eqnarray}
where the integer $\omega_j= y'_{s-1} +(j-z_s)y'_{s}$ if $z_s \leq j <z_{s+1}$, $j=1, 2, \cdots, z_l$, and $\omega_{z_l+1}= y'_l$ if $j=z_l+1$.

From Eq.(\ref{pf}), we have $0 \leq (-1)^s q_j \leq p_s$ if $z_s \leq j < z_{s+1}$, which leads to
\begin{eqnarray}\label{qjp}
0 \leq (-1)^s(\omega_j - \frac{n_j}{p_0} ) \leq \frac{p_{s}}{p_0}.
\end{eqnarray}
The value of $\omega_j$ depends on the parity of $s$. If $s$ is even, $\omega_j$ can be expressed as
\begin{eqnarray}\label{wjf1}
\omega_j= \left\lfloor \frac{n_j}{p_0} \right\rfloor +1, \quad z_s \leq j < z_{s+1},
\end{eqnarray}
Based on Eqs.(\ref{qjp}) and (\ref{wjf1}), we construct the inequality
\begin{eqnarray}
\left\lfloor \frac{n_j}{p_0} \right\rfloor +1 -\frac{p_s +1}{p_0} \leq \frac{n_j-1}{p_0}  \leq \left\lfloor \frac{n_j}{p_0} \right\rfloor +1 -\frac{1}{p_0}. \label{qreven2}
\end{eqnarray}
The relation (\ref{p11f}) with even $s\geq 2$ gives
\begin{eqnarray}\label{qreven3}
 0\leq 1 -\frac{p_s +1}{p_0} <1.
\end{eqnarray}
Taking the integer part of Eq.(\ref{qreven2}) and using Eq.(\ref{qreven3}), we obtain
\begin{eqnarray}
  \left\lfloor\frac{n_j-1}{p_0} \right\rfloor   = \left\lfloor \frac{n_j}{p_0} \right\rfloor .
\end{eqnarray}
If $s=0$, due to the fact $1 \leq n_j < p_0$ if $ 1 \leq j < z_1$, we have
\begin{eqnarray}
  \left\lfloor \frac{n_j-1}{p_0} \right\rfloor   = \left\lfloor \frac{n_j}{p_0} \right\rfloor=0, \qquad  1 \leq j < z_1.
\end{eqnarray}
From the above discussions, we arrive at
\begin{eqnarray}
\omega_j=\lfloor \frac{n_j-1}{p_0} \rfloor +1.\label{qrodwed4}
\end{eqnarray}

If $s$ is odd, Eq.(\ref{qjp}) tells us that
\begin{eqnarray}\label{wjodd2}
  \omega_j =\left\lfloor \frac{n_j}{p_0} \right\rfloor.
\end{eqnarray}
Based on (\ref{qjp}) and (\ref{wjodd2}), we construct the inequality
\begin{eqnarray}
 \left\lfloor \frac{n_j}{p_0} \right\rfloor - \frac{1}{p_0} \leq \frac{n_j-1}{p_0}  \leq \left\lfloor \frac{n_j}{p_0} \right\rfloor +\frac{p_s -1}{p_0}.\label{qrodd2}
\end{eqnarray}
The relation (\ref{p11f}) with odd $s\geq 3$ gives
\begin{eqnarray}\label{qrodd3}
-1< \frac{p_s -1}{p_0}< 0.
\end{eqnarray}
Taking the integer part of Eq.(\ref{qrodd2}) and using (\ref{qrodd3}), we obtain
\begin{eqnarray}
 \left\lfloor \frac{n_j-1}{p_0} \right\rfloor = \left\lfloor \frac{n_j}{p_0} \right\rfloor -1.\label{qrosddd4-1}
\end{eqnarray}
If $s=1$ and $ z_1 < j < z_2$, based on Eqs.(\ref{qjp}) and (\ref{wjodd2}), we construct another inequality
\begin{eqnarray}\label{qrodd4}
\left\lfloor \frac{n_j}{p_0} \right\rfloor -\frac{1 }{p_0} \leq \frac{n_j-1}{p_0} < \left\lfloor \frac{n_j}{p_0} \right\rfloor +\frac{p_1 -1}{p_0}.
\end{eqnarray}
Taking the integer part of Eq.(\ref{qrodd4}) and considering
\begin{eqnarray}\label{qrodd5}
\frac{p_1 -1}{p_0}= 0,
\end{eqnarray}
we obtain
\begin{eqnarray}
\left\lfloor \frac{n_j-1}{p_0} \right\rfloor = \left\lfloor \frac{n_j}{p_0} \right\rfloor -1.\label{qrosddd4-2}
\end{eqnarray}
If $s=1$ and $j=z_1$, the $\omega_{z_1}=0$ because of $q_{z_1}=-1$.
Comparing (\ref{wjodd2}), (\ref{qrosddd4-1}) and (\ref{qrosddd4-2}), we find that $ \omega_j$ can also be written as (\ref{qrodwed4}). Then we arrive at
the conclusion (\ref{qwr2}).

\section*{Appendix C: Result for the periodic XYZ model}
\setcounter{equation}{0}
\label{sec-com}
\setcounter{equation}{0}
\renewcommand{\theequation}{C.\arabic{equation}}

\subsection*{C.1 The system}

In order to study the effect induced by twisted boundaries, we consider the XYZ spin chain (\ref{Hamiltoniant}) with periodic boundary condition (c.f. (\ref{Antiperiodic}))
\begin{eqnarray}
\sigma^{\alpha}_{N+1} =\sigma^{\alpha}_1, \quad \alpha=x,y,z.
\end{eqnarray}
In this case, the eigenvalue of the Hamiltonian (\ref{Hamiltoniant}) reads
\begin{eqnarray}\label{eigEp}
  E_p(\eta)=\frac{\sigma(\eta)}{\sigma'(0)} \left\{ \sum^M_{j=1} \left[ \frac{\sigma'(\nu_j)}{\sigma(\nu_j)} -\frac{\sigma'(\mu_j +\eta)}{\sigma(\mu_j +\eta)} \right] +\frac{N}{2}\frac{\sigma'(\eta)}{\sigma(\eta)} + 2i \pi l_1 \right\},
\end{eqnarray}
where the $2M$ Bethe roots $\{ \mu_j \}$ and  $\{ \nu_j \}$ should satisfy the BAEs
\begin{eqnarray}
  && (\frac{N}{2}-M)\eta -\sum^M_{j=1}(\mu_j-\nu_j) = l_1  \tau +m_1, \quad l_1,m_1 \in \textbf{Z}, \nonumber  \\
  && M\eta +\sum^M_{j=1}(\mu_j +\nu_j) =  m_2, \quad m_2\in \textbf{Z},\nonumber  \\
  &&  c_p e^{[ 2i\pi l_1(\mu_j +\eta)\mu_j  +i\phi_p ]}\sigma^{\bar L_1}(\mu_j+\frac{\eta}{2}) \sigma^N(\mu_j +\eta) = \prod^M_{l=1}\sigma(\mu_j-\nu_l) \sigma(\mu_j-\nu_l+ \eta), \nonumber \\
  && c_p e^{-2i\pi l_1 \nu_j -i\phi_p } \sigma^{\bar L_1} (\nu_j +\frac{\eta}{2}) \sigma^N(\nu_j) =- \prod^M_{l=1}\sigma(\nu_j-\mu_l) \sigma(\nu_j-\mu_l- \eta),\nonumber  \\
  && e^{i\phi_p}\prod^M_{j=1} \frac{\sigma(\mu_j+\eta)}{\sigma(\nu_j)}=e^{\frac{2i\pi k_1}{N}}, \quad k_1 =1,\cdots, N. \label{batp2}
\end{eqnarray}
Here $c_p$ and $\phi_p$ are the constants, and the non-negative integer $\bar L_1$ satisfies the constraint
\begin{eqnarray}
  N+\bar L_1=2M.
\end{eqnarray}

The BAEs (\ref{batp2}) are derived from the inhomogeneous $T-Q$ relation \cite{cao2014spin}. If the system size $N$ is even,
detailed analysis of Eq.(\ref{batp2}) shows that either $\mu_j=\nu_l$ or
$\mu_j=\nu_l-\eta$ leads to $c_p=0$ and hence induces a one-to-one correspondence
between Bethe roots $\{\mu_j\}$ and $\{\nu_l\}$, which means only one set of Bethe roots is survived. Therefore, for a generic $\eta$, if we require
\begin{eqnarray}
l_1=0, \quad N=2M, \quad \{\mu_j\}=\{\nu_l\},
\end{eqnarray}
then $c_p=0$ and the BAEs reduce to the conventional ones, which are
consistent with the results given in references \cite{baxter2000partition,baxter1972one,takhtadzhan1979quantum}.
That is the reason why the periodic XYZ model with even $N$ has been studied extensively.
However, if the system size $N$ is odd, this conclusion is not true for a generic $\eta$. The correct statement is that only
at following discrete points of crossing parameter $\eta$
\begin{eqnarray}\label{etap}
  \bar{\eta}_{m_1,l_1}=\frac{2l_1}{N-2M}\tau +\frac{2m_1}{N-2M},\quad l_1,m_1\in \textbf{Z},
\end{eqnarray}
the parameter $c_p=0$ thus the BAEs (\ref{batp2}) can reduce to the conventional ones.
Here we consider the case that $\eta$ is real, which can be achieved by putting $l_1=0$.
Without losing generality, we also put $M=N$ and $m_1=-m$. Then the degenerate points (\ref{etap}) becomes
\begin{eqnarray}\label{etap2}
  \bar{\eta}_m=\frac{2m}{N},\quad m\in \textbf{Z},
\end{eqnarray}
and the reduced BAEs are
\begin{eqnarray}
  && e^{2i\phi_p} \frac{\sigma^N [\frac{i}{2}(x_j-\bar{\eta}_m i)] }{\sigma^N [\frac{i}{2}(x_j+\bar{\eta}_m i)] }
  =\prod^N_{k\neq j} \frac{\sigma [\frac{i}{2}(x_j-x_k -2\bar{\eta}_m i)] }{\sigma [\frac{i}{2}(x_j-x_k +2\bar{\eta}_m i)]},\quad j=1,\cdots, N, \nonumber \\
  &&e^{i\phi_p} \prod^N_{j=1} \frac{\sigma [\frac{i}{2}(x_j-\bar{\eta}_m i)] }{\sigma [\frac{i}{2}(x_j +\bar{\eta}_m i)] }=e^{\frac{2i\pi k_1}{N}},
  \quad k_1 =1,\cdots, N, \label{BAEp22}
\end{eqnarray}
where $\{ x_j \}$ are the Bethe roots in the degenerate cases. Accordingly, we obtain the eigenvalue of system at the degenerate points as
\begin{eqnarray}\label{eigEp2}
  E_{p}(\bar \eta_m) =\frac{\sigma(\bar{\eta}_m)}{\sigma'(0)} \left\{ \sum^N_{j=1} \left[ \frac{\sigma' [\frac{i}{2}(x_j +\bar{\eta}_m i)] }{\sigma [\frac{i}{2}(x_j +\bar{\eta}_m i)] } -\frac{\sigma' [\frac{i}{2}(x_j -\bar{\eta}_m i) ]}{\sigma [\frac{i}{2}(x_j -\bar{\eta}_m i)]} \right] +\frac{N}{2}\frac{\sigma'(\bar{\eta}_m)}{\sigma(\bar{\eta}_m)} \right\}.
\end{eqnarray}

\begin{figure}[!htp]
 \centering
    \includegraphics[height=6cm]{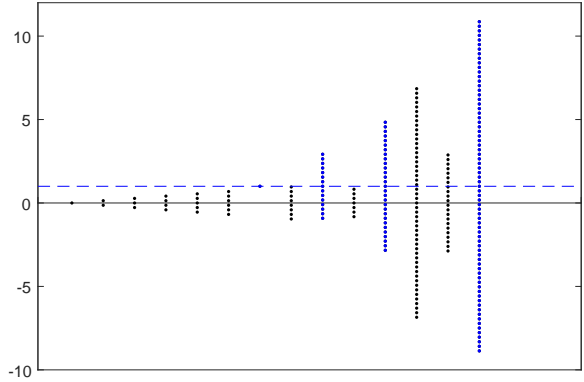}
\caption{
Strings on the complex plane for $\bar \eta_m=\frac{2m}{N}=\frac{10}{73}$.
The strings are arranged from left to right in the order of string length determined by Eq.(\ref{str-rule}).
The blue dashed line is given by shifting the real axis up by unit. We see the strings are symmetric distribute around the real axis or the blue dashed line.
}\label{Esadg-oddp}
\end{figure}

Using the same idea as suggested previously, we first solve the reduced BAEs (\ref{BAEp22}) and
obtain the exact results at the degenerate points $\bar \eta_m$. Then we extrapolate these results to the real $\eta$ case.
Here, we only list the main results and neglect all the detailed treatment.
Due to the fact that $\bar \eta_m$ is real, in the present case, the strings in the complex plane are parallel to the imaginary axis and are
symmetric around the real axis or the $y=i$ line, which is shown in Fig.\ref{Esadg-oddp}.
We consider the same region $0<\bar \eta_m \leq \frac{1}{2}$ as before.

\subsection*{C.2 Ground state}

Repeating the similar calculation with antiperiodic case, we obtain the ground state energy of XYZ model with odd $N$ as
\begin{eqnarray}\label{Egoddpr}
  E^{odd}_{p,g}(\eta)=e_0(\eta) N +\epsilon_h(\frac{\tau}{i},\eta).
\end{eqnarray}
Here $e_0(\eta)$ is the ground state energy density which is given by Eq.(\ref{Eden}) after replacing the $\eta_m$ by $\eta$.
$\epsilon_h(x^h, \eta)$ is the energy of hole at the point of $x^h$, which can be obtained by replacing $\eta_m$ in Eq.(\ref{Eh3}) with $\eta$.
Obviously, the function $\epsilon_h(x^h, \eta)$ takes its minimum value if $x^h=\tau/i$.

Comparing Eqs.(\ref{Egoddpr}) and (\ref{Egoddr}), we find that the ground state energy of XYZ model with periodic boundary condition equals to
that with antiperiodic one. It is reasonable because that there is only one bound is twisted in Hamiltonian (\ref{Hamiltoniant}). In the thermodynamic limit,
it is hard that the twisted bound can affect the energy of whole system. However, the twisted bound can affect the eigenstates.
The eigenstates of periodic and that of antiperiodic are totally different.

\begin{figure}[!htp]
    \centering
    \includegraphics[height=4.5cm]{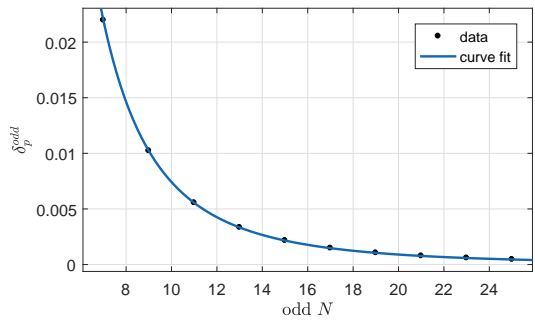}
    \caption{
The deviation $\delta^{odd}_{p}$ versus the system-size $N$. The data can be fitted as $ \delta ^{odd}_{p}=\alpha_5 N^{\beta_5}$.
Here $\tau =0.5i$, $\eta=\frac{\sqrt{2}}{4}$, $\alpha_5=8.225$ and $\beta_5=-3.045$.
Due to the fact $\beta_5< 0$, the deviation $\delta^{odd}_{p} \rightarrow  0$ when $N\rightarrow \infty$.}\label{Eg-oddp}
\end{figure}

Now, we check the formula (\ref{Egoddpr}) by the exact diagonalization. Define
\begin{eqnarray}\label{dEg-oddp}
  \delta^{odd}_{p}= \frac{E^{odd}_{p,g}(\eta)}{\bar{E}^{odd}_{p,g}(\eta)}-1,
\end{eqnarray}
where $ E^{odd}_{p,g}(\eta)$ is the ground state energy obtained by Eq.(\ref{Egoddpr})
and $\bar{E}^{odd}_{p,g}(\eta)$ is the ground state energy obtained by the exact diagonalization.
In Table \ref{table-oddp}, we list the numerical results with $\tau=0.5i$ and $\eta =0.4$. We find that
the deviation $\delta^{odd}_{p}$ is about $10^{-4}$ if $N=25$,
which means that Eq.(\ref{Egoddpr}) is still valid for the finite system size provided that $N$ is not too small.
In Fig.\ref{Eg-oddp}, the $\delta^{odd}_{p}$ for various $N$
with $\tau=0.5i$ and $\eta=\sqrt{2}/4$ are shown.
The data can be fitted as power law
\begin{eqnarray}
  \delta^{odd}_{p}=\alpha_5 N^{\beta_5}.
\end{eqnarray}
Due to the fact $\beta_5< 0$, the value $\delta^{odd}_{p}$ tends to zero when the system size $N \rightarrow \infty$,
which indicates that Eq.(\ref{Egoddpr}) can describe the ground state energy in the thermodynamic limit.
\begin{table}[!htp]
  \caption{
  The numerical results of periodic XYZ model with $\eta=0.4$ and $\tau=0.5i$. Here,
  $\bar{E}^{odd}_{p,g}$ is the ground state energy obtained by the exact diagonalization,
  $E^{odd}_{p,g}$ is the ground state energy obtained by Eq.(\ref{Egoddpr}),
  and $\delta ^{odd}_{p}$ is the deviation.
  }
  \centering
\begin{tabular}{| r|r|r|r|}\hline
$ N$ & $  \bar{E}^{odd}_{p,g}$ & $  E^{odd}_{p,g}$ & $ \delta^{odd}_{p}~(\times 10^{-2})$  \\ \hline
 $ 5 $ & $-4.0730$  & $-4.2577$  & $4.5353$   \\ \hline
 $ 7 $ & $-6.4903$  & $-6.5909$  & $1.5507$   \\ \hline
 $ 9 $ & $-8.8616$  & $-8.9241$  & $0.7057$   \\ \hline
 $ 11 $ & $-11.2149$  & $-11.2573$  & $0.3779$   \\ \hline
 $ 13 $ & $-13.5600$  & $-13.5905$  & $0.2253$   \\ \hline
 $ 15 $ & $-15.9007$  & $-15.9237$  & $0.1449$   \\ \hline
 $ 17 $ & $-18.2389$  & $-18.2569$  & $0.0986$   \\ \hline
 $ 19 $ & $-20.5757$  & $-20.5901$  & $0.0701$   \\ \hline
 $ 21 $ & $-22.9115$  & $-22.9233$  & $0.0516$   \\ \hline
 $ 23 $ & $-25.2467$  & $-25.2565$  & $0.0391$   \\ \hline
 $ 25 $ & $-27.5814$  & $-27.5897$  & $0.0303$   \\ \hline
 \end{tabular}
\label{table-oddp}
\end{table}

The ground state energy with even $N$ is
\begin{eqnarray}\label{Egevenpr}
  E^{even}_{p,g}(\eta)=e_0(\eta) N.
\end{eqnarray}
Comparing Eqs.(\ref{Egevenpr}) and (\ref{Egevenr}) , we find that the ground state energy with periodic boundary condition
and that with antiperiodic one are the same.

\subsection*{C.3 Elementary excitation}

\begin{figure}[!htp]
    \centering
    \includegraphics[height=4.5cm]{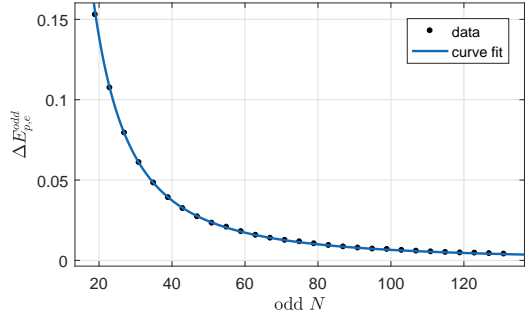}
    \caption{
The energy of hole excitation $\Delta E^{odd}_{p,e}$ versus the system-size $N$.
The data can be fitted as $ \Delta E^{even}_{p,e}=\alpha_6 N^{\beta_6}$.
Here $\tau =0.5i$, $\eta=\frac{\sqrt{2}}{4}$, $\alpha_6=41.18$ and $\beta_6=-1.899$.
Due to the fact $\beta_6< 0$, the $\Delta E^{odd}_e$ tends to zero when $N \rightarrow \infty$.}\label{exc-odd-p}
\end{figure}

Repeating the similar calculation with antiperiodic case, we obtain
the energy of hole excitation with odd $N$
\begin{eqnarray}\label{Ecoddrp}
  \Delta E^{odd}_p (x^h, \eta)= \epsilon_h(x^h,\eta) -\epsilon_h(\frac{\tau}{i},\eta).
\end{eqnarray}
In the thermodynamic limit, the position of hole $x^h$ can tend to $\tau/i$ infinitely, thus
\begin{eqnarray}\label{Efoddp}
\Delta E^{odd}_{p}(\eta) = \lim_{x^h\rightarrow \frac{\tau}{i}} \Delta E^{odd}_p (x^h,\eta) \rightarrow 0.
\end{eqnarray}
Now, we check the analytical result (\ref{Efoddp}) by the DMRG method. The numerical results, denoted as $ \Delta E^{odd}_{p,e}$, with $\tau=0.5 i$ and $\eta=\sqrt{2}/4$ are shown in Fig.\ref{exc-odd-p}.
The data are fitted as
\begin{eqnarray}
  \Delta E^{odd}_{p,e} =\alpha_6 N^{\beta_6}.
\end{eqnarray}
Due to the fact $\beta_6< 0$, the $\Delta E^{odd}_{p,e}$ tends to zero when the system size $N \rightarrow \infty$, which is consistent with Eq.(\ref{Efoddp}).
Comparing Eqs.(\ref{Efoddp}) and (\ref{Efodd}), we also find that the excitation energy with periodic boundary condition
is the same as that with antiperiodic one.

\begin{figure}[!htp]
    \centering
    \includegraphics[height=4.5cm]{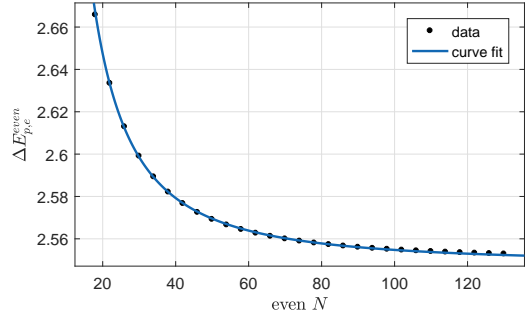}
    \caption{
The energy of hole excitation $\Delta E^{even}_{p,e}$ versus the system-size $N$.
The data can be fitted as $ \Delta E^{even}_{p,e}=\alpha_7 N^{\beta_7} +\epsilon_7$.
Here $\tau =0.5i$, $\eta=\frac{\sqrt{2}}{4}$, $\alpha_7=14.71$, $\beta_7=-1.668$ and $\epsilon_7=2.548$.
Due to the fact $\beta_7< 0$, in the thermodynamic limit, $\epsilon_7$ gives the gap, i.e., $\epsilon_7= \Delta E^{even}_{p}(\eta)$.}\label{exc-even-p}
\end{figure}

The hole excitation with even $N$ has a gap
\begin{eqnarray}\label{Efevenp}
\Delta E^{even}_{p}(\eta)=2\epsilon_h(\frac{\tau}{i},\eta).
\end{eqnarray}
Comparing Eqs.(\ref{Efevenp}) and (\ref{Efeven}),
we find that the gap with periodic boundary condition is the same as that with antiperiodic one.
Now, we check the analytic result (\ref{Efevenp}) by the DMRG method.
The numerical results, denoted as $\Delta E^{even}_{p,e}$, with $\tau=0.5 i$ and $\eta=\sqrt{2}/4$ are given in Fig.\ref{exc-even-p}.
The data can be fitted as
\begin{eqnarray}
  \Delta E^{even}_{p,e}=\alpha_7 N^{\beta_7} +\epsilon_7. \label{Eferevenp}
\end{eqnarray}
Due to the fact $\beta_7<0$, in the thermodynamic limit, $\epsilon_7$ gives the energy gap.
The DMRG data give $\epsilon_7= 2.548$, which is highly consistent with the analytical results (\ref{areven}). We note that although the values of fitting factors $\alpha_7$ and $\beta_7$
in (\ref{Eferevenp}) are different from that of $\alpha_4$ and $\beta_4$ in (\ref{aresdven}), the energy gap $\epsilon_7=\epsilon_4$.

Finally, let us remark that by taking the limit $\tau \rightarrow i \infty$ of above results, one can easily obtain corresponding results for the periodic XXZ spin chain.

\end{document}